\documentclass[%
reprint, 
groupedaddress,
 amsmath,amssymb,
 amsfonts,
 aps,
 prd,
floatfix
]{revtex4-2}

\usepackage{graphicx}
\usepackage{dcolumn}
\usepackage{bm}
\usepackage{hyperref} 
\usepackage{xcolor}
\usepackage[section]{placeins}
\usepackage{multirow}

\hypersetup{linktocpage=true } 

\begin{document}


\title{Microphysics of early dark energy}

\author{Vivian I. Sabla}
\email{vivian.i.sabla.gr@dartmouth.edu}
\author{Robert R. Caldwell}%
\affiliation{%
 Department of Physics and Astronomy, Dartmouth College, \\ HB 6127 Wilder Laboratory, Hanover, NH 03755 USA
}%

\date{\today}

\begin{abstract}
Early dark energy (EDE) relies on scalar field dynamics to resolve the Hubble tension, by boosting the pre-recombination length scales and thereby raising the CMB-inferred value of the Hubble constant into agreement with late universe probes. However, the collateral effect of scalar field microphysics on the linear perturbation spectra appears to preclude a fully satisfactory solution. $H_0$ is not raised without the inclusion of a late universe prior, and the ``$S_8$ tension'', a discrepancy between early- and late-universe measurements of the structure growth parameter, is exacerbated. What if EDE is not a scalar field? Here, we investigate whether different microphysics, encoded in the constitutive relationships between pressure and energy density fluctuations, can relieve these tensions. We show that EDE with an anisotropic sound speed can soften both the $H_0$ and $S_8$ tensions while still providing a quality fit to CMB data. Future observations from the CMB-S4 experiment may be able to distinguish the underlying microphysics at the $4\sigma$ level, and thereby test whether a scalar field or some richer physics is at work.
\end{abstract}

\maketitle

\section{Introduction}

The $\Lambda$-cold dark matter ($\Lambda$CDM) model has become the standard model of cosmology due to its success at describing a vast range of cosmological measurements, from the cosmic microwave background (CMB) \cite{Planck:2018vyg}, to large-scale structure (LSS) \cite{BOSS:2016wmc,DES:2021wwk}, to the expansion history of the Universe \cite{Riess:2020fzl}. However, the increased precision of cosmological probes has led to tensions arising between these measurements. Most notably, early-universe inferences of the present-day cosmic expansion rate using the CMB are consistently and significantly lower than late- (or local-) universe measurements of $H_0$. This so-called ``Hubble tension'' can be seen most clearly between the SH0ES measurement of $H_0=73.04\pm1.04$ km/s/Mpc \cite{Riess:2021jrx} and the \textit{Planck} inference of $H_0=67.4\pm 0.5$ km/s/Mpc \cite{Planck:2018vyg}, which differ by more than $5\sigma$. Albeit, some late-time measurements are in agreement with both the CMB inferred values of $H_0$ and other late-time measurements, within uncertainties \cite{Pesce:2020xfe,Huang:2019yhh,Freedman:2019jwv}. But as it stands, the Hubble tension does not seem to be able to be explained away by systematics in any experiment \cite{Efstathiou:2013via,Addison:2015wyg,Planck:2016tof,Aylor:2018drw}. This discrepancy between model-independent, local measurements and $\Lambda$CDM-based, early inferences of $H_0$ suggests that there may be new physics beyond the $\Lambda$CDM model at play, particularly in the pre-recombination era \cite{Bernal:2016gxb,Verde:2016wmz,Freedman:2017yms,Feeney:2017sgx,Evslin:2017qdn,Verde:2019ivm,Knox:2019rjx}. The many attempts to resolve the Hubble tension  fall in two broad categories: late-universe solutions which alter local-universe determinations of $H_0$, and early-universe solutions which change pre-recombination physics to alter the CMB inference of $H_0$ (see \cite{DiValentino:2021izs,Schoneberg:2021qvd} and references therein).
 
Early dark energy (EDE) has proven to be one of the most promising classes of early-universe solutions, with many models significantly reducing the $H_0$ tension, while yielding a comparable fit to the observational data compared to $\Lambda$CDM \cite{Poulin:2018dzj,Poulin:2018cxd,Smith:2019ihp,Murgia:2020ryi,Smith:2020rxx,Niedermann:2019olb,Niedermann:2020dwg,Niedermann:2020qbw,Freese:2021rjq,Sakstein:2019fmf,CarrilloGonzalez:2020oac,Agrawal:2019lmo,Lin:2019qug,Hill:2021yec,Poulin:2021bjr,Karwal:2021vpk,Ye:2021iwa}. A leading example is standard EDE, consisting of a scalar field that briefly bumps up the expansion rate between equality and recombination.

However, the perturbative dynamics of the canonical scalar field used in these models appear to preclude a fully satisfactory solution. \textit{Planck} data alone does not favor EDE as a cosmological model. It is only with the inclusion of a late-universe prior on $H_0$ that EDE is favored in non-negligible amounts. This is avoided in analyses with alternative CMB datasets, yet more work needs to be done to determine if these differing constraints are physical, or due to experimental systematics \cite{Poulin:2021bjr, Hill:2021yec,Smith:2022hwi}. 
More importantly, EDE models tend to exacerbate the discrepancy between early- and late-universe measurements of the structure growth parameter, known as the ``$S_8$ tension'' \cite{Hill:2020osr,Ivanov:2020ril,DAmico:2020ods,DiValentino:2020vvd,Clark:2021hlo,Cai:2021weh,Vagnozzi:2021gjh,Sabla:2021nfy,Nunes:2021ipq}. A possible interpretation of this continuing tension is that new physics is required, but it may not be a canonical scalar field.

Previous studies have investigated the implications of varying the sound speed in EDE-like models from its canonical scalar field value with favorable results \cite{Smith:2019ihp,Lin:2019qug}. In this work, we pursue a description of EDE as a phenomenological fluid component, whose background evolution is matched to a family of viable EDE models. The perturbative dynamics of this EDE fluid are specified by a gauge-invariant sound speed, relating pressure and density perturbations, and a gauge-invariant anisotropic stress, modeled either via an equation of state inspired by proposed dark sector stress models \cite{Battye:2007aa,Battye:2013aaa}, or via an equation of motion formalism inspired by generalized dark matter models \cite{Hu:1998kj}. We refer to the constitutive relations necessary to define this perturbative sector as the microphysics of EDE. We explore what types of perturbative evolution is necessary to strengthen and improve on current EDE solutions to the Hubble tension, without specifying a particular physical model.

For canonical EDE scalar fields, the microphysics is fixed: the evolution of the perturbations follow that of a single, uncoupled fluid with no anisotropic shear, and a relationship between the density and pressure giving a gauge-invariant sound speed of $c_\phi^2=1$. By exploring deviations from this framework, specifically with the addition of anisotropic shear, we implicitly explore the viability of non-scalar field EDE. Examples of theoretical models that yield anisotropic stress include free-streaming neutrinos \cite{Weinberg:2003ur} as well as more exotic cases such as topological defects, cosmic lattice models such as elastic dark energy, and coherent vector fields \cite{Bucher:1998mh,Battye:1999eq,Battye:2005hw,Battye:2005mm,Battye:2013aaa,Battye:2014xna, Soergel:2014sna,Bielefeld:2015daa}. However, these models are not specifically known to predict the full range of properties (equation of state history, parallel and perpendicular sound speeds) that are investigated in this paper. Instead, we have constructed phenomenological models with generalized properties beyond the specific examples. This approach is similar in spirit to generalized dark matter \cite{Hu:1998kj} and elastic dark energy \cite{Battye:2007aa,Battye:2013aaa} constructions.

We find that EDE with an added anisotropic stress can simultaneously soften both the $H_0$ and $S_8$ tensions when compared to a scalar field EDE model, implying that EDE need not be the result of a scalar field. Current data cannot definitively discriminate among these possibilities, but future measurements may be more decisive.

This paper is organized as follows. In Sec.~\ref{sec: PFM} we outline our phenomenological fluid parametrization of EDE. We describe the background solution to the Hubble tension as well as the different microphysics variations that we analyze. In Sec.~\ref{sec: data}, we present the cosmological data used to test the viability of the different microphysics scenarios. Results are given in Sec. \ref{sec: results}, and we conclude our discussion in Sec. \ref{sec: conclusions} with our main findings. Additional details about the models considered and extended results can be found in the Appendixes.

\section{Phenomenological Fluid Model (PFM)}
\label{sec: PFM}

We work in a scenario consisting of the standard cosmological model with cold dark matter (CDM), and dark energy in the form of a cosmological constant. We introduce an EDE component in the form of a phenomenological fluid with perturbative dynamics that differ from a canonical scalar field. In this section we describe the background fluid dynamics and introduce the constitutive relations necessary to consistently describe the microphysics. 

\subsection{Background dynamics}
\label{sec: background}
\begin{figure}[t]
     \centering
     \includegraphics[width=\linewidth]{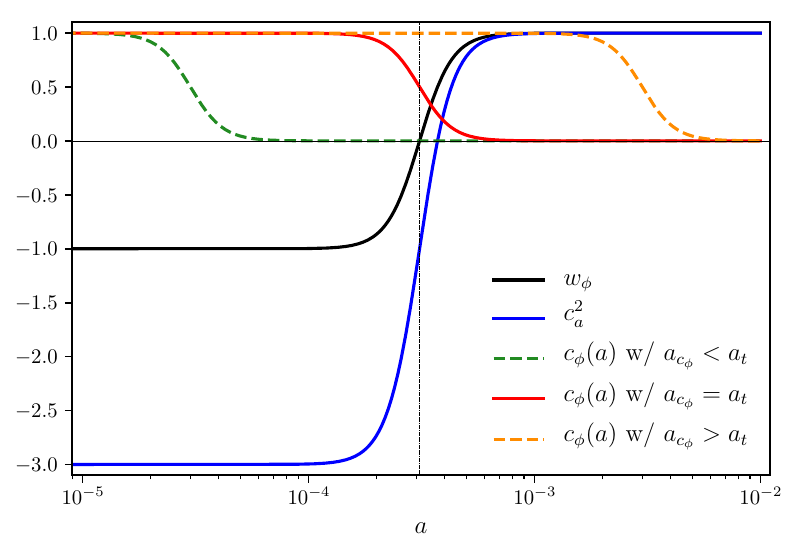}
     \caption{The evolution of the equation of state $w_\phi$, the adiabatic sound speed $c_a^2$, and the dynamical sound speed $c_\phi(a)$ for different values of $a_{c_\phi}$ as a function of scale factor for a model with $n=6$ and $a_t=a_{c_\phi}=3.1\times 10^{-4}$. The vertical black dotted line delineates the transition scale factor $a_t$.}
     \label{fig: ca2}
 \end{figure}

\begin{figure}[t]
     \centering
     \includegraphics[width=\linewidth]{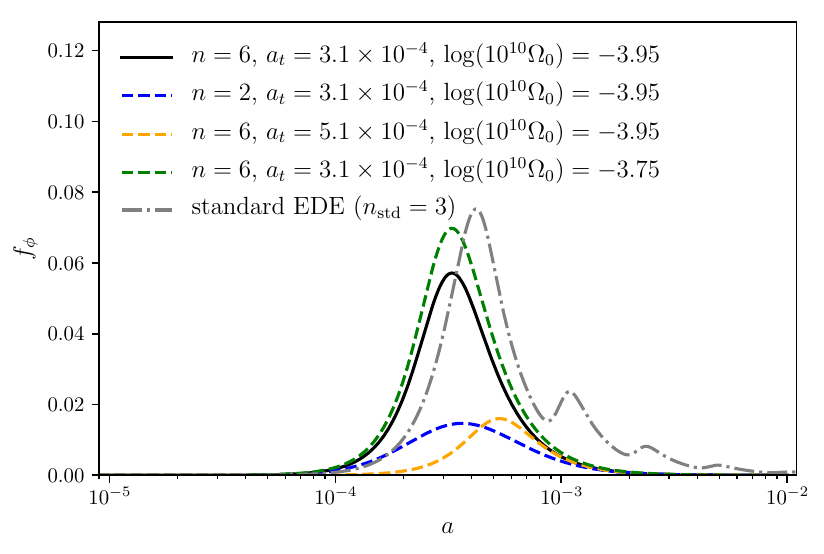}
     \caption{The evolution of $f_\phi=\rho_\text{tot,$\phi$}/\rho_\text{tot,$\Lambda$CDM} -1$ as a function of scale factor. The black solid line shows our baseline model with $n=6$, $\log(10^{10}\Omega_0)=-3.95$, and $a_t=3.1 \times 10^{-4}$. The blue dashed line shows the baseline model with $n=2$. The orange dashed line shows the baseline model with $a_t=5.1 \times 10^{-4}$. The dotted green line shows the baseline with $\log(10^{10}\Omega_0)=-3.75$. The gray dot-dashed line shows the best-fit $n_\text{std}=3$ oscillating scalar field model of EDE from Ref.~\cite{Smith:2019ihp} for comparison. }
     \label{fig: fede}
 \end{figure}

Our proposed scenario mimics the background evolution of standard EDE by specifying a time-varying equation of state
\begin{equation}
    w_\phi(a) = -1 + \frac{2}{1+(a_t/a)^n}, 
    \label{eq: wphi}
\end{equation}
which evolves from $w_\phi=-1$ to $w_\phi=1$ at a time given by the transition scale factor $a_t$, as shown in Fig.~\ref{fig: ca2}. The sharpness of this transition is controlled by the parameter $n$ with higher $n$ corresponding to a faster and sharper transition in the equation of state. This transition in $w_\phi$ resembles the thaw of a scalar field from the Hubble friction, causing the energy density in our phenomenological fluid to spike in a similar fashion, as shown in Fig. \ref{fig: fede}. We control the amount of energy added by this fluid component by setting the energy density of EDE in the present day $\rho_{\phi,0}$. In turn, the evolution of the energy density in the phenomenological fluid is given by
\begin{equation}
    \rho_\phi(a) = \rho_{\phi,0} a^{-6} \left(\frac{1+a_t^n}{1+(a_t/a)^n} \right)^{6/n}.
    \label{eq: rhophi}
\end{equation}
Equations (\ref{eq: wphi}) and (\ref{eq: rhophi}) give us a background model for EDE as a phenomenological fluid, constituting a 3-parameter extension to $\Lambda$CDM.

Our EDE fluid model differs from the effective fluid approximation of EDE presented in Ref.~\cite{Poulin:2018cxd} mainly through the definition of the equation of state. Reference \cite{Poulin:2018cxd} defines
\begin{equation}
    w_\phi(a) = \frac{1+w_n}{1+(a_c/a)^{3(1+w_n)}}-1, 
\end{equation}
where $w_n=(n_\text{std}-1)/(n_\text{std}+1)$. Here the ``std'' subscript indicates Ref.~\cite{Poulin:2018cxd} variables as distinct from our parameters. In Ref.~\cite{Poulin:2018cxd}, the equation of state transitions from a value of $w_\phi=-1$ to $w_n$ at a time specified by a transition scale factor $a_c$. Hence, the parameter $n_\text{std}$ effectively controls the rate at which the energy density in the field dilutes after becoming dynamical. In contrast, the final equation of state in our phenomenological EDE fluid is fixed to $w_\phi=1$, with the energy density diluting as $a^{-6}$, and our parameter $n$ controls the sharpness of the transition from the starting value of $w_\phi$. For constraints on the final equation of state see Refs.~\cite{Poulin:2018cxd,Lin:2019qug}. We choose the parametrization given in Eq.~(\ref{eq: wphi}) to more generally model a component whose energy density ``gets out of the way'' fast enough to not have adverse effects at the background level, allowing us to focus on the perturbative changes discussed in Sec.~\ref{sec: microphysics}. However, these two parametrizations are equivalent for the cases of $n_\text{std}=\infty$ and $n=6$.
At the background level, with $n=6$, $a_t=3.1\times10^{-4}$, and $\log(10^{10}\Omega_0)=-3.95$, where $\Omega_0=\rho_{\phi,0}/\rho_\text{crit,0}$, this model faithfully reproduces the standard EDE best-fit $n_\text{std}=\infty$ model of Ref.~\cite{Poulin:2018cxd}.  

EDE as a solution to the Hubble tension is grounded in the theoretical description of the CMB angular power spectrum. The CMB is sensitive to $H_0$ via the angular size of the first acoustic peak, which can be modeled as $\theta_s=r_s(z_*)/D_A(z_*)$, where $r_s(z_*)$ is the comoving sound horizon at decoupling, and $D_A(z_*)$ is the comoving angular diameter distance to the surface of last scattering. The sound horizon, which depends on pre-recombination background energy densities, scales as $H_0^{-1/2}$, whereas $D_A(z_*)$, which is dependent on post-recombination energy densities, scales as $H_0^{-1}$. We may decrease the size of the sound horizon by adding new components to the energy density, and thereby expect an increase in the value of $H_0$ inferred from the CMB.

The above-described procedure is based solely on the background cosmological model. However, linear perturbations of all components of the cosmic fluid play a significant role in the creation of the CMB angular power spectrum. 

\subsection{Perturbative microphysics}
\label{sec: microphysics}

Standard, scalar field EDE is relativistic and does not cluster, which is manifest in the behavior of the density and velocity perturbations. As a result, EDE perturbations lead to an enhancement in power of the first acoustic peak of the CMB when compared to \textit{Planck} data, which is compensated for by an increase in the dark matter density $\omega_{cdm}$, and a subsequent increase in the spectral index $n_s$. These changes lead to a larger $S_8$, increasing the $S_8$ tension, and restricting the amount of EDE allowed by the data. In fact, standard EDE never fully resolves the $H_0$ tension. In this work, we frame our EDE model as a phenomenological fluid and examine how changes to the microphysics affect the clustering response, and ultimately the proposed solution to the $H_0$ tension. 

For a generalized fluid component, the standard description of linearized perturbations requires the specification of four variables: energy density, pressure, momentum density, and anisotropic stress. Traditionally, the pressure perturbation $\delta p_\phi$ is set via $c_s^2=\delta p_\phi/\delta\rho_\phi$, where $c_s^2$ is the effective sound speed, normally set equal to unity, and $\delta\rho_\phi$ is the density perturbation. However, this formulation is gauge dependent and lacks generality, so to be as exhaustive as possible we consider a gauge-invariant formulation of the pressure perturbation 
\begin{equation}
\label{eq: pressure pert}
    \delta p_\phi = c_\phi^2 \delta\rho_\phi +3 \mathcal{H} (c_\phi^2-c_a^2)(\rho_\phi+p_\phi) \theta_\phi/k^2,
\end{equation}
where we work in Fourier space, $\theta_\phi$ is the velocity divergence, and $c_\phi^2$ is now the effective sound speed of our fluid, which is a free parameter in this generalized fluid formulation. The adiabatic sound speed $c_a^2$ takes on a simple form in our phenomenological model
\begin{equation}
\label{eq: ca2 def}
    c_a^2\equiv \frac{p_\phi'}{\rho_\phi'}=w_\phi-\frac{w_\phi'}{3\mathcal{H}(1+w_\phi)}= w_\phi -\frac{n}{6}(1-w_\phi),
\end{equation}
where primes denotes derivatives with respect to conformal time $'=\partial{}/\partial{\tau}$. In our scenarios, $w_\phi$ starts near $-1$ at early times. Consequently, the adiabatic sound speed starts out negative before evolving towards $c_a^2=1$ on the same time scale as the transition in the equation of state, as shown in Fig.~\ref{fig: ca2} for a case with $n=6$. The higher the value of $n$, the more negative the starting value of $c_a^2$. 

The conservation of the stress-energy tensor yields two equations of motion for the density contrast of the fluid $\delta_\phi=\delta\rho_\phi/\rho_\phi$ and the velocity divergence,
\begin{multline}
\label{eq: delta prime}
    \delta_\phi' = -3\mathcal{H}(c_\phi^2-w_\phi)\delta_\phi-(1+w_\phi)\frac{h'}{2}\\ -\left[ k^2 +9\mathcal{H}^2(c_\phi^2-c_a^2) \right](1+w_\phi)\theta_\phi/k^2
\end{multline}
\begin{equation}
\label{eq: theta prime}
    \theta_\phi' = -\mathcal{H}(1-3c_\phi^2)\theta_\phi +\frac{c_\phi^2}{1+w_\phi} k^2\delta_\phi  - k^2\sigma_\phi
\end{equation}
where $\mathcal{H}=a'/a$ is the conformal Hubble parameter, $\sigma_\phi$ is the anisotropic stress, $h$ is the synchronous gauge metric potential (see \cite{Ma:1995ey}), and we have explicitly used our definition of the gauge-invariant pressure perturbation. From Eq.~(\ref{eq: delta prime}) and (\ref{eq: theta prime}) we can see there are two free parameters that define the evolution of linear perturbations: the effective sound speed of the fluid, $c_\phi^2$, and the anisotropic shear $\sigma_\phi$. It is through variation of these parameters that we test noncanonical EDE microphysics.

\subsubsection{Varied sound speed}
\begin{table}[t]
\begin{tabular}{| c | c | c | c |}
\hline \hline
 Case & $c_\phi^2$ & $A_\sigma$ & Description \\ \hline \hline
 1 & 0 & 0 & pressure-less fluid \\ 
 2 & $c_\phi(a)$ & 0 &  dynamic $c_\phi$ with $a_{c_\phi}=a_t$\\
 3 & $c_\phi(a)$ & 0 & dynamic $c_\phi$ with  $a_{c_\phi}<a_t$\\
 4 & $c_\phi(a)$ & 0 & dynamic $c_\phi$ with  $a_{c_\phi}>a_t$\\
 \hline \hline 
\end{tabular}
\caption{\label{tab: summary of models} Outline of various sound speed cases considered for the shear-less PFM model. In case 1 the sound speed is constant and set to $c_\phi^2=0$. Cases 2-4 consider the dynamical sound speed presented in Eq.~(\ref{eq: dyn c_phi}) with different transition scale factors.}
\end{table}
Our first test of microphysics comes with shear-less models where the only perturbative parameter we have to define is the sound speed $c_\phi^2$. The role of the gauge-invariant sound speed is dependent upon the time rate of change of the background equation of state. For a slowly varying equation of state, $c_\phi^2$ determines the fluctuation response on subhorizon scales. Here, slow means $w_\phi’/w_\phi \lesssim a’/a$. Whereas for a rapidly varying equation of state, $w_\phi’/w_\phi \gtrsim a’/a$, a new scale is introduced into the system. Roughly speaking, for subhorizon perturbations in the range $a’/a \lesssim k  \lesssim w_\phi’/w_\phi$, the effective sound speed may differ dramatically. Consider $c_\text{eff}^2 = \langle \delta p \rangle / \langle \delta\rho\rangle$, where the angle brackets indicate an appropriate time averaging. Only on smaller scales, $k \gtrsim w_\phi’/w_\phi$, does the gauge-invariant sound speed $c_\phi^2$ play an important role. It is in this way that a rapidly oscillating scalar field can achieve a nonrelativistic sound speed, despite $c_\phi^2=1$. (See, for example, Refs.~\cite{Turner:1983he,Cembranos:2015oya}.) 

For the phenomenological fluid model described in Sec.~\ref{sec: background}, the equation of state changes slowly, allowing $c_\phi^2=1$ to represent a canonical scalar field. It would certainly be possible to introduce rapid variations in $w_\phi$ that affect the microphysics. However, designing such a $w_\phi(t)$ time history seems baroque in the absence of a particular underlying model to serve as a guide. Hence, for the purposes of this work, a fluid with $c_\phi^2=1$ will be used to delineate a scalar field-like model, with deviations from this value probing alternative microphysics scenarios. 

We consider two distinct scenarios with varied sound speeds: (i) a constant sound speed $c_\phi^2=0$, giving a pressureless fluid, and (ii) a dynamical formulation of the sound speed such that 
\begin{equation}
\label{eq: dyn c_phi}
    c_\phi^2(a) = 1 - \frac{1}{1+(a_{c_\phi}/a)^n}. 
\end{equation}
In this dynamical model, the sound speed will transition from $c_\phi^2=1$, to $c_\phi^2=0$ at a time specified by a critical scale factor $a_{c_\phi}$, as shown in Fig.~\ref{fig: ca2}. Similarly to the transition in the background equation of state, the sharpness of this transition is set by $n$. This transition in the sound speed can either occur simultaneously with the transition in the equation of state such that $a_{c_\phi}=a_t$, or $a_{c_\phi}$ can be altered to happen before or after the background transition as shown in Fig.~\ref{fig: ca2}. In this way we have four distinct cases of non-canonical sound speeds which we outline in Table \ref{tab: summary of models}. Case 1 gives our pressureless fluid with a constant sound speed $c_\phi^2=0$. Cases 2-4 consider the dynamical sound speed model with different values of sound speed transition scale factor $a_{c_\phi}$.

The acoustic dark energy (ADE) model of Ref.~\cite{Lin:2019qug} explores the phenomenology of noncanonical constant sound speed in a similar EDE fluid model. They find that joint variations to the final equation of state and sound speed of the fluid can improve on the fit to cosmological data given in a canonical case. Our variations to the sound speed differ in that the final equation of state is held fixed at $w_\phi=1$, and the sound speed is allowed to vary independently from the background evolution of the fluid. In this way, we explicitly probe changes to the perturbative sector, decoupled from the background fluid dynamics. 

The background dynamics of our fluid give $w_\phi$, and $c_\phi^2$ is a free parameter that can vary as described above, leaving the anisotropic shear as the only variable left to define. Standard EDE has no shear, so we must look to other components for realistic models. We consider two different shear models, described below.

\subsubsection{Shear models}
\label{sec: shear models}
For our first shear model, we follow the approach suggested in Ref.~\cite{Battye:2013aaa} to model dark sector stress in terms of a gauge invariant equation of state and define
\begin{equation}
\label{eq: shear model 1}
    (1+w_\phi)\sigma_\phi = A_\sigma \left[\delta_\phi+3\mathcal{H}(1+w_\phi)\theta_\phi/k^2\right],
\end{equation}
where $A_\sigma$ is a free scaling parameter. Depending on the sign of $A_\sigma$ this shear is built to damp or enhance the growth of the velocity perturbation at late times, resulting in changes to the evolution of the EDE density perturbation at the same scales. We will henceforth refer to this equation-of-state shear model, Eq.~(\ref{eq: shear model 1}), as shear model I. 

The second shear model we consider is derived from the density and velocity perturbations of our generalized fluid component and defines an equation of motion for the shear
\begin{equation}
\label{eq: shear model 2}
    \sigma_\phi'+3\mathcal{H}(c_\phi^2-c_a^2)\sigma_\phi + A_\sigma(\theta_\phi+\alpha k^2)=0,
\end{equation}
where $\alpha=(h'+6\eta')/2k^2$, and $A_\sigma$ is again a free scaling parameter, just like in our previous shear model. Similarly to the generalized dark matter (GDM) stress presented in Ref.~\cite{Hu:1998kj}, we choose the shear to be sourced by the velocity perturbation $\theta_\phi$, with the metric perturbation term included for gauge invariance. In the limit that $c_\phi^2=1$ and $w_\phi=0$, this equation exactly matches the GDM stress of Ref.~\cite{Hu:1998kj}. By setting $c_\phi^2=w_\phi=1/3$, we recover the equation of motion for the stress given by a Boltzmann hierarchy for radiation truncated at the quadrupole \cite{Ma:1995ey}.
This equation of motion formalism of the shear, Eq.~(\ref{eq: shear model 2}), will henceforth be referred to as shear model II. Details on the specific form of both shear models considered can be found in Appendix \ref{sec: shear model details}.

\subsubsection{Generalized shear behavior}
To better understand the physical meaning of our microphysics parameters $c_\phi^2$ and $A_\sigma$, it is instructive to think about the anisotropic shear within the context of the stress energy of our fluid. By perturbing the stress-energy tensor it is simple to show that shear is only nonzero when the pressure response to some scalar perturbation is anisotropic, with $\delta p= (\delta p_x+\delta p_y+\delta p_z)/3$
and $(\rho+p)\sigma =-(\hat k_i \hat k_j - \frac{1}{3} \delta_{ij})\delta T^i_j$.  Let us now consider a mode travelling in the $\hat z$ direction, and rotate our system such that $\delta p_x=\delta p_y=\delta p_\perp$ making $\delta p_z=\delta p_\parallel$. Combining this framework with our two shear models we find that we can write $c_\phi^2$ and $A_\sigma$ in terms of the perpendicular and parallel sound speeds of the fluid, such that
\begin{equation}
\label{eq: cphi2}
    c_\phi^2 = \frac{1}{3}\left(2 \frac{\delta p_\perp}{\delta \rho}+ \frac{\delta p_\parallel}{\delta\rho}\right) = \frac{1}{3}(2 c_\perp^2+c_\parallel^2), 
\end{equation}
and 
\begin{equation}
\label{eq: Asig}
    A_\sigma =\frac{2}{3}\left( \frac{\delta p_\perp}{\delta\rho} - \frac{\delta p_\parallel}{\delta\rho} \right)= \frac{2}{3}(c_\perp^2-c_\parallel^2),
\end{equation}
in both shear models we consider. As expected, $c_\phi^2$ is just the spatially averaged sound speed of our fluid. The amount of shear in our fluid, parametrized by $A_\sigma$, is controlled by the difference in the directional sound speeds. For an isotropic pressure perturbation, $A_\sigma=0$, and there is no shear contribution. When the pressure perturbation becomes anisotropic, $A_\sigma \neq 0$ and we have a nonzero shear contribution. This framework also gives us physical bounds on our model parameters $c_\phi^2$ and $A_\sigma$. For stability and causality, $0 \leq c_\perp^2,c_\parallel^2 \leq 1$, which restricts $0 \leq c_\phi^2 \leq 1$ and $-2/3\leq A_\sigma \leq 2/3$. 

The equations of motion given in Eq.~(\ref{eq: delta prime}) and Eq.~(\ref{eq: theta prime}), coupled with either stress model constitute a full description of the perturbative sector dynamics. Besides the three background parameters $n$, $\Omega_0$, and $a_t$, there are now two additional free parameters describing the microphysics of our phenomenological EDE fluid, $c_\phi^2$ and $A_\sigma$.

\section{Data and Methodology}
\label{sec: data}
We derive cosmological parameter constraints by running a complete Markov Chain Monte Carlo (MCMC) using the public code \texttt{CosmoMC} (see \url{https://cosmologist.info/cosmomc/}) \cite{Lewis:2002ah} with modified versions of the Boltzmann solver \texttt{CAMB} to solve the different linearized perturbations in each of our microphysics scenarios \cite{Lewis:1999bs}. We model neutrinos as two massless and one massive species with $m_\nu=0.06$ eV and $N_\text{eff}=3.046$. Our dataset includes different combinations of early and late time data described below:
\begin{itemize}
    \item \textbf{P18:} \textit{Planck} 2018 CMB measurements via the \texttt{TTTEEE plik lite} high-$\ell$, \texttt{TT} and \texttt{EE} low-$\ell$, and lensing likelihoods \cite{Planck:2019nip}.
    \item \textbf{BAO:} We use data from the BOSS survey (data release 12) at $z=$0.38, 0.51, and 0.61 \cite{BOSS:2016wmc}, low redshift measurements from the 6dF survey at $z=$ 0.106 \cite{Beutler:2011hx}, and the BOSS main galaxy sample at $z=0.15$ \cite{Ross:2014qpa}.
    \item \textbf{R19:} Local Hubble constant measurement by the SH0ES collaboration giving $H_0=74.03\pm1.42$ km/s/Mpc \cite{Riess:2020fzl}.
    \item \textbf{SN:} Pantheon supernovae dataset consisting of the luminosity distances of 1048 SNe Ia in the redshift range of $0.01<z<2.3$ \cite{Scolnic:2017caz}. 
\end{itemize}
Note that the \texttt{plik lite} likelihood is a foreground and nuisance marginalized version of the full \texttt{plik} likelihood \cite{Planck:2019nip}. We have found that the two likelihoods return nearly identical posterior distributions with statistically equivalent $\Delta\chi^2$ values for cases of standard EDE as a phenomenological fluid, as well as cases with altered sound speed and nonzero anisotropic stress. Therefore, we use the \texttt{plik lite} likelihood in place of the full likelihood for speed in analysis. 

We start our analysis of the full phenomenological fluid model by setting our microphysics to match a canonical scalar field model with $c_\phi^2=1$ and $A_\sigma=0$ and obtain constraints on our background model parameters, as well as the six standard $\Lambda$CDM parameters \{$\omega_b,\omega_c,\theta_s,\tau,\ln(10^{10} A_s),n_s$\}. This serves as a proof of concept that this phenomenological EDE fluid model can resolve the Hubble tension in an equivalent manner to scalar field EDE models. We then fix the background model parameters, and vary the standard model parameters along with only our added microphysics parameters to directly test the impact of the altered microphysics scenarios. We first set $A_\sigma=0$ and consider four variations to the gauge-invariant sound speed $c_\phi^2$, outlined in Table \ref{tab: summary of models}. We then incorporate the two shear models presented in Sec.~\ref{sec: shear models} and parametrize our microphysics via Eqs.~(\ref{eq: cphi2})-(\ref{eq: Asig}) to derive constraints on our microphysics parameters $c_\phi^2$ and $A_\sigma$. When considering models with anisotropic shear, we assume the sound speed of the fluid is constant, but allowed to vary from the canonical value of $c_\phi^2=1$. The dynamical $c_\phi^2(a)$ model given by Eq.~(\ref{eq: dyn c_phi}) is only used in models with no anisotropic shear (i.e. $A_\sigma=0$).  We perform our analysis with a Metropolis-Hasting algorithm with flat priors on all parameters. Our results were obtained by running eight chains and monitoring convergence via the Gelman-Rubin criterion, with $R-1<0.05$ for all parameters considered complete convergence \cite{Gelman:1992zz}.

\section{Results}
\label{sec: results}

\begin{table*}[t]
\begin{center}
\begin{tabular}{| c | c | c | c | c |}
\hline \hline
  & \multicolumn{2}{c|}{P18 only} & \multicolumn{2}{c|}{P18+BAO+R19+SN}\\ 
  \hline
 Parameter & $\Lambda$CDM & PFM &  $\Lambda$CDM  & PFM \\ \hline \hline
 $100 \omega_b$ & $2.235(2.237)\pm 0.015$ & $2.241(2.248)^{+0.017}_{-0.022}$ & $2.252(2.250)\pm0.013$ & $2.287(2.291)\pm0.022$\\ 
 $\omega_c$ & $0.1202(0.1199)\pm 0.0013$ & $0.1227(0.1224)^{+0.0017}_{-0.0029}$ & $0.11830(0.11843)\pm0.00091$ & $0.1269(0.1283)^{+0.0031}_{-0.0034}$\\ 
 $100 \theta_s$ & $1.04089(1.04105)\pm0.00032$ & $1.04071(1.04081)\pm0.00034$ & $1.04115(1.04106)\pm0.00028$ & $1.04063(1.04061)\pm0.00035$\\
 $\tau$ & $0.0553(0.0551)\pm0.0076$ & $0.0541(0.0573)^{+0.0070}_{-0.0079}$ & $0.0608(0.0584)^{+0.0072}_{-0.0081}$ & $0.0575(0.0541)\pm0.0074$\\ 
 $\ln (10^{10} A_s)$ & $3.046(3.045)\pm0.015$ & $3.049(3.056)\pm0.015$ & $3.055(3.051)^{+0.014}_{-0.016}$ & $3.065(3.063)\pm0.015$\\ 
 $n_s$ & $0.9645(0.9644)\pm0.0043$ & $0.9668(0.9714)^{+0.0046}_{-0.0061}$ & $0.9693(0.9685)\pm0.0038$ & $0.9809(0.9844)\pm0.0065$ \\
 $1/n$ & - & $<0.525(0.136)$ & - & $<0.249(0.159)$\\
 $r_\phi$ & - & $<0.0222(0.0242)$ & - & $0.071(0.090)^{+0.027}_{-0.030}$ \\
 $a_t \times 10^{4}$ & - & $4.02(2.78)^{+0.17}_{-1.50}$ & - & $3.07(3.12)^{+0.22}_{-0.44}$ \\
 \hline
 $H_0$ [km/s/Mpc] & $67.27(67.45)\pm0.56$ & $67.90(68.07)^{+0.63}_{-0.91}$ & $68.16(68.07)\pm0.41$ & $70.32(70.82)\pm0.89$ \\ 
 $S_8$ & $0.834(0.829)\pm0.013$ & $0.840(0.841)\pm0.014$ & $0.816(0.817)\pm0.010$ & $0.840(0.841)\pm0.013$\\
 \hline
 Total $\chi^2_\text{min}$ & 1014.09 & 1012.72 & 2073.37 & 2063.33 \\
 $\Delta \chi^2_\text{min}$ & - & -1.37 & - & -10.04 \\
 \hline \hline 
\end{tabular}
\end{center}
\caption{\label{tab: background constraints} The mean (best-fit) $\pm 1\sigma$ error of the cosmological parameters for $\Lambda$CDM and the PFM with $c_\phi^2=1$ and $A_\sigma=0$ for the P18 dataset and a combined P18+BAO+R19+SN dataset. For the PFM the microphysics parameters are held fixed and constraints are derived on the background model parameters only. }
\end{table*}

In the following section we explore the implications of this phenomenological EDE fluid model on CMB-derived cosmological parameters. We start by holding the microphysics parameters fixed to their canonical values, and show that our fluid model gives a comparable resolution to the Hubble tension at the background level to standard EDE \cite{Smith:2019ihp,Poulin:2018cxd}. Next, we vary only the effective sound speed of the fluid and show that altering $c_\phi^2$ alone does not improve the standard EDE solution. 
We then evaluate our two shear models presented in Sec.~\ref{sec: shear models} and find that shear model II with $c_\phi^2 \sim 0.55$ and $A_\sigma \sim -0.2$ not only improves the resolution to the Hubble tension provided by standard EDE, but also softens the $S_8$ tension in comparison to the standard EDE solution, all while providing as comparable a fit to \textit{Planck} 2018 data as the $\Lambda$CDM model. Finally, we use Fisher forecasting to determine the power of CMB-S4 \cite{CMB-S4:2016ple} to distinguish altered microphysics from the standard EDE case. Extended results can be found in Appendix \ref{sec: extended results}. 

\subsection{Resolution to the Hubble tension}
\label{sec: resolution to the H0 tension}

 \begin{figure*}[t]
     \centering
     \includegraphics[height=0.9\textheight]{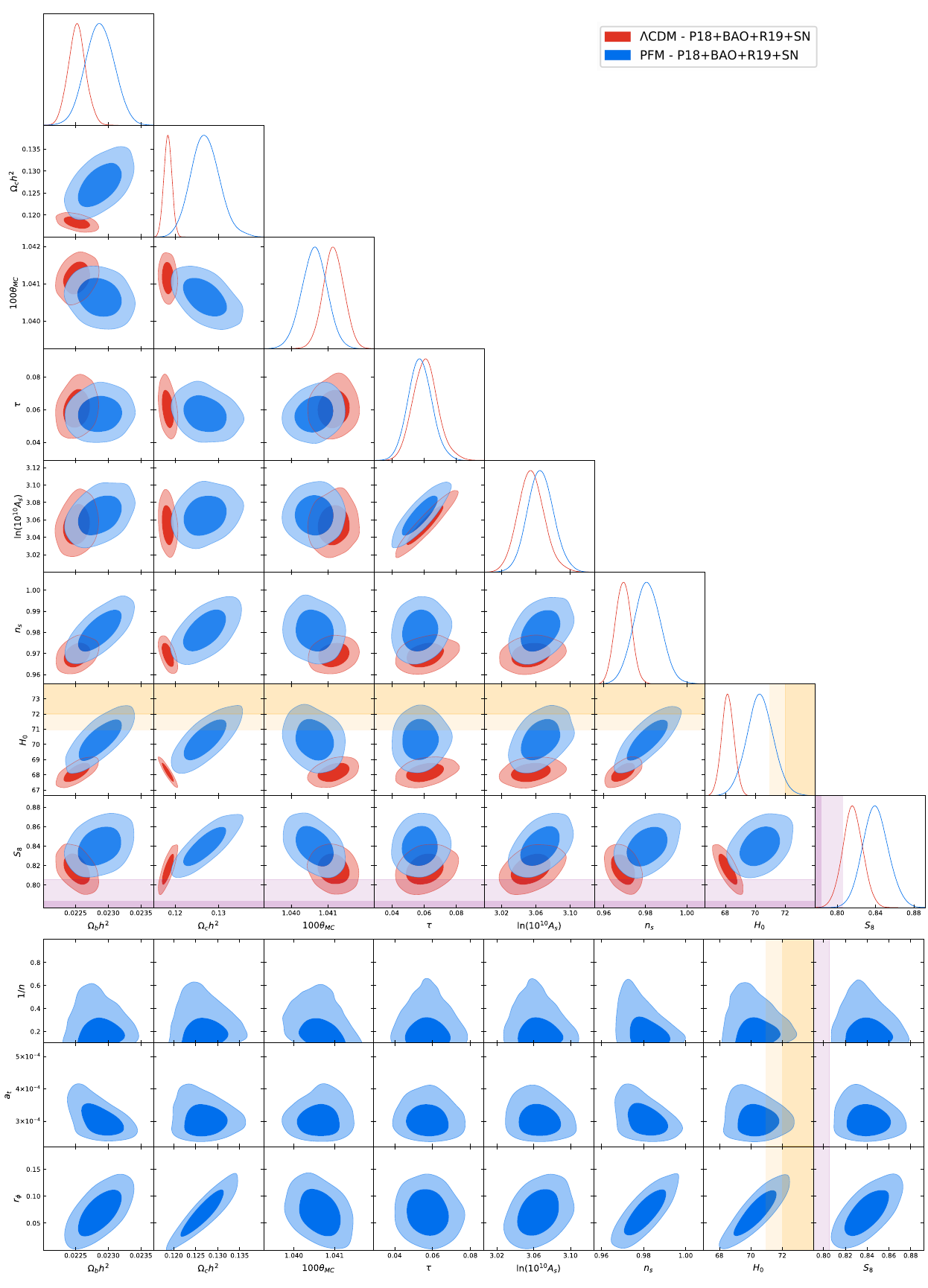}
     \caption{
     Top: Posterior distributions of the standard model parameters for the $\Lambda$CDM model (red) and the PFM with $c_\phi^2=1$ and $A_\sigma=0$ (blue). Bottom: Posterior distributions of the standard model parameters vs. the background PFM parameters for the PFM with $c_\phi^2=1$ and $A_\sigma=0$. 
     The darker inner (lighter outer) regions correspond to $1\sigma$($2\sigma$) confidence intervals. The SH0ES collaboration measurement of $H_0=73.04\pm1.04$ km/s/Mpc and the KiDS-1000 weak lensing survey measurement of $S_8=0.759^{+0.024}_{-0.021}$ are shown in the orange and purple bands, respectively \cite{Riess:2021jrx,KiDS:2020suj}. Distributions are generated with the P18+BAO+R19+SN datasets.
     }
     \label{fig: background constraints}
 \end{figure*}
 \begin{figure}[t]
     \centering
     \includegraphics[width=\linewidth]{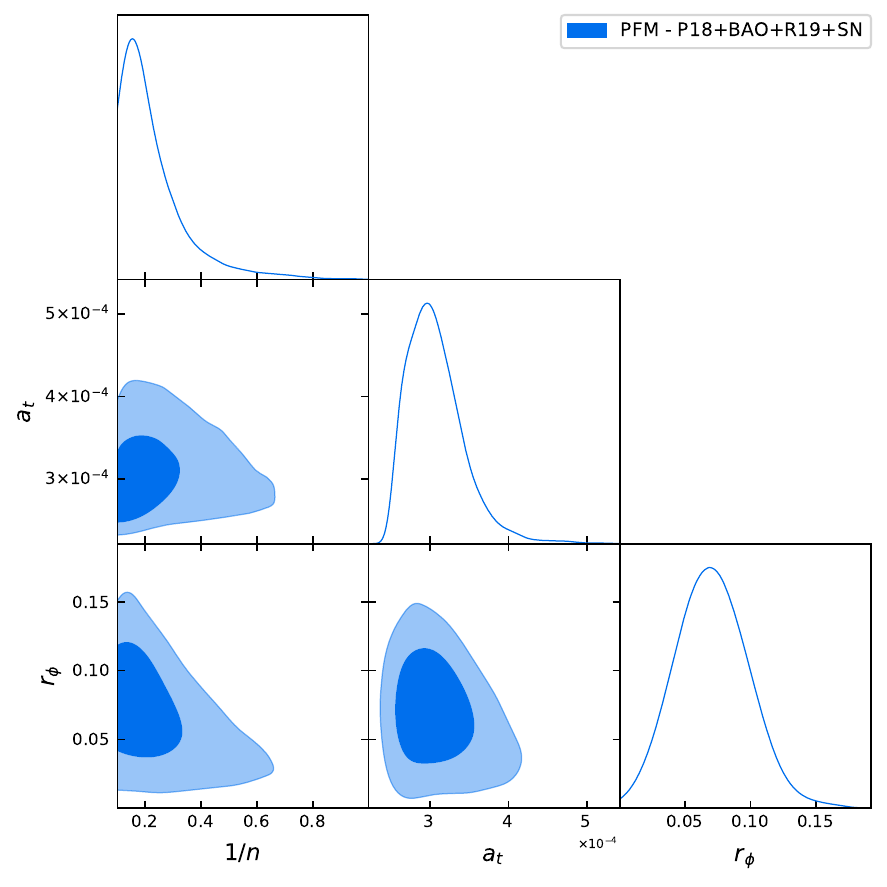}
     \caption{Posterior distributions of the background PFM parameters for the PFM with $c_\phi^2=1$ and $A_\sigma=0$. The darker inner (lighter outer) regions correspond to $1\sigma$($2\sigma$) confidence intervals. Distributions are generated with the P18+BAO+R19+SN datasets.
     }
     \label{fig: background constraints model params}
 \end{figure}

We begin our analysis by confirming that our phenomenological fluid model can resolve the Hubble tension in the same way as standard EDE. We set the microphysics of our fluid such that it behaves like a canonical scalar field with $c_\phi^2=1$ and $A_\sigma=0$. In order to facilitate convergence of the MCMC chains, we reparametrize our background model and derive constraints on $r_\phi\equiv \rho_\phi(a_t)/\rho_{\Lambda\text{CDM}}(a_t)$, where
$\rho_{\Lambda\text{CDM}}(a_t)$ represents the contribution to the energy density of the standard model components at the transition scale factor $a_t$. Furthermore, instead of varying the sharpness parameter $n$, we derive constraints on $1/n$ for computational ease. This background parametrization in terms of $r_\phi$ is distinct from, but analogous to, the effective-fluid approximation given in Ref.~\cite{Poulin:2018cxd} where the fractional density is set via $f_\text{ede}(z_c)\equiv\Omega_\phi(z_c)/\Omega_\text{tot}(z_c)$ with $z_c$ defining the time at which the fluid becomes dynamical. We assume flat priors on all parameters, with $0.1<1/n<1$, $0<r_\phi<1$, and $2.49< a_t \times 10^4<9.21$, to keep the transition before recombination. 
We show parameter constraints for the background model and standard model parameters in Table \ref{tab: background constraints} for the PFM with a combination of different datasets. The best-fit $\Lambda$CDM model is shown for comparison. Posterior distributions for all relevant parameters are shown in Figs.~\ref{fig: background constraints} and \ref{fig: background constraints model params}. 

As can be seen in Table \ref{tab: background constraints}, our phenomenological fluid provides a similar resolution to the Hubble tension as the standard EDE model of Refs.~\cite{Poulin:2018cxd,Smith:2019ihp}, when considering the same datasets. For a combined analysis using the P18+BAO+R19+SN datasets outlined in Sec.~\ref{sec: data}, this fluid model of EDE yields a best-fit value of $H_0=70.82$ km/s/Mpc, reducing the Hubble tension with late-universe measurements to $\sim 2\sigma$, while fitting the full suite of data better than $\Lambda$CDM with $\Delta\chi^2_\text{min}=-10.4$. We find a preference for a nonzero amount of this EDE fluid at $2\sigma$ with $r_\phi=0.071^{+0.027}_{-0.030}$, peaking at $a_t=3.07^{+0.22}_{-0.44}\times 10^{-4}$. As expected, the $S_8$ tension is exacerbated with $S_8=0.840\pm0.013$. Similarly to scalar field EDE, this appreciable increase in $H_0$ only occurs when a late-universe prior on the Hubble constant is used in analysis. Considering \textit{Planck} 2018 data alone yields $H_0=68.07$ km/s/Mpc, in statistical agreement with the best-fit $\Lambda$CDM value. Putting this all together, this phenomenological fluid model proves to behave just like a standard EDE model at the background level. 

Furthermore, these results show good agreement with the ADE model presented in Ref.~\cite{Lin:2019qug}. While there is no direct parameter mapping between our two models, our phenomenological fluid EDE model is comparable to the canonical ADE model of Ref.~\cite{Lin:2019qug}, as we consider the same microphysics model with different parametrizations of a phenomenological standard EDE fluid. Our constraints on the full dataset shown in Table \ref{tab: background constraints} are in good agreement with the cADE constraints given in Table I of Ref.~\cite{Lin:2019qug}. The variations between our constraints can be explained by slight differences in our models and analysis. Our analysis of our phenomenological fluid EDE model considers an extra parameter, $n$, when compared to Ref.~\cite{Lin:2019qug}, however we still achieve similar results. Furthermore, we base our analysis on the updated \textit{Planck} 2018 data \cite{Planck:2018vyg} as opposed to the \textit{Planck} 2015 data \cite{Planck:2015fie}. Despite these differences, our results are still statistically comparable to the cADE parametrization, suggesting consistency between our two models. 

With these results we have shown that we can mimic the solution to the Hubble tension that scalar field EDE provides without specifying a particular physical model. In order to directly compare the impact of altering the microphysics in this model, we define a baseline case for which the background model parameters are fixed to $n=6$, $\log(10^{10}\Omega_0)=-3.95$, and $a_t=3.1\times 10^{-4}$. For this baseline case, the microphysics parameters are also held fixed at their canonical values of $c_\phi^2=1$ and $A_\sigma=0$. This baseline model gives a 5.7\% spike in the energy density right around matter-radiation equality, matching the background evolution of the best-fit $n_\text{std}=\infty$ model of Ref.~\cite{Poulin:2018cxd}. Parameter constraints for this baseline model are given in Table \ref{tab: baseline P18}. 

We can now use this phenomenological fluid model to assess the viability of nonscalar field EDE via altered microphysics. Specifically, the presence of anisotropic shear can be used as a diagnostic of any EDE models that arise from anisotropic media, like cosmic strings or cosmic lattice models \cite{Bucher:1998mh,Battye:2005mm} or coherent vector fields \cite{Bielefeld:2015daa}, where isotropy is preserved at the background level, but broken in the evolution of linear perturbations. These deviations from the baseline case manifest as changes to the microphysics of our phenomenological fluid, which we discuss in the following sections. 

\begin{table}[t]
\begin{tabular}{| c | c | c |}
\hline \hline
 Parameter & PFM - baseline case \\ \hline \hline
 $100 \omega_b$ & $2.263(2.263)\pm0.015$ \\ 
 $\omega_c$ & $0.1261(0.1259)\pm0.0011$ \\ 
 $100 \theta_s$ & $1.04059(1.04065)\pm0.00029$ \\
 $\tau$ & $0.0542(0.0564)\pm0.0072$ \\ 
 $\ln (10^{10} A_s)$ & $3.057(3.061)\pm0.014$\\
 $n_s$ & $0.9747(0.9760)\pm0.0042$ \\
 $n$ & 6 (fixed) \\
 $\log(10^{10} \Omega_0)$ & -3.95 (fixed) \\
 $a_t$ & 0.00031 (fixed) \\
 \hline
 $H_0$ [km/s/Mpc] & $69.04(69.11)\pm0.58$ \\ 
 $S_8$ & $0.849(0.849)\pm0.013$ \\
 \hline
 Total $\chi^2_\text{min}$ & 1013.39 \\
 $\Delta \chi^2_\text{min}$ & -0.70 \\
 \hline \hline 
\end{tabular}
\caption{\label{tab: baseline P18} The mean (best-fit) $\pm 1\sigma$ error of the cosmological parameters for baseline PFM with $c_\phi^2=1$ and $A_\sigma=0$. Constraints are based on the P18 dataset. }
\end{table}

\subsection{Varying the sound speed}
\label{sec: results cs}

\begin{table*}[t]
\begin{tabular}{| c | c | c | c | c |}
\hline \hline
 Parameter & PFM - case 1  & PFM - case 2 & PFM - case 3 & PFM - case 4\\ \hline \hline
 $100 \omega_b$ & $2.185(2.176)\pm0.014$ & $2.274(2.280)\pm0.015$ & $2.184(2.182)\pm0.014$ & $2.263(2.266)\pm0.015$\\ 
 $\omega_c$ & $0.1111(0.1117)\pm0.0014$ & $0.1163(0.1160)\pm0.0013$ & $0.1112(0.1114)\pm0.0014$ & $0.1261(0.1264)\pm0.0012$\\ 
 $100 \theta_s$  & $1.03974(1.03966)\pm0.00030$ & $1.04081(1.04082)\pm0.00031$ & $1.03975(1.03966)\pm0.00030$ & $1.04059(1.04063)\pm0.00030$\\
 $\tau$ & $0.0642(0.0622)^{+0.0072}_{-0.0093}$ & $0.0641(0.0650)^{+0.0074}_{-0.0091}$ & $0.0636(0.0619)\pm0.0082$ & $0.0541(0.0528)\pm0.0077$\\ 
 $\ln (10^{10} A_s)$ & $3.053(3.049)^{+0.014}_{-0.018}$ & $3.070(3.071)^{+0.014}_{-0.017}$ & $3.051(3.047)\pm0.016$ & $3.057(3.056)\pm0.015$\\ 
 $n_s$ & $0.9617(0.9602)\pm0.0042$ & $0.9707(0.9713)\pm0.0044$ & $0.9615(0.9613)\pm0.0041$ & $0.9746(0.9733)\pm0.0043$\\
 $a_{c_\phi}\times 10^4$ & - & 3.1 (fixed) & 0.3 (fixed) & 30 (fixed) \\
 \hline
 $H_0$ [km/s/Mpc] & $74.39(73.97)\pm0.79$ & $73.27(73.45)\pm0.73$ & $74.34(74.13)\pm0.80$ & $69.02(68.96)\pm0.60$\\ 
 $S_8$ & $0.743(0.749)\pm0.014$ & $0.752(0.749)\pm0.013$ & $0.742(0.744)\pm0.015$ & $0.849(0.850)\pm0.013$\\
 \hline
 Total $\chi^2_\text{min}$ & 1326.47 & 1087.30 & 1323.68 & 1013.28\\
 $\Delta \chi^2_\text{min}$ & +312.38 & +73.21 & +309.59 & -0.81 \\
 \hline \hline 
\end{tabular}
\caption{\label{tab: vary sound speed P18} The mean (best-fit) $\pm 1\sigma$ error of the cosmological parameters for cases 1-4 of our PFM model with varied sound speeds, outlined in Table \ref{tab: shear II P18}. The background model is fixed for all cases to $n=6$, $\log(10^{10}\Omega_0)=-3.95$, and $a_t=3.1\times 10^{-4}$ for direct comparison to the baseline case. All cases considered have no added anisotropic shear. Constraints are based on the P18 dataset. }
\end{table*}
 \begin{figure*}[t]
     \centering
     \includegraphics[width=0.955\textwidth]{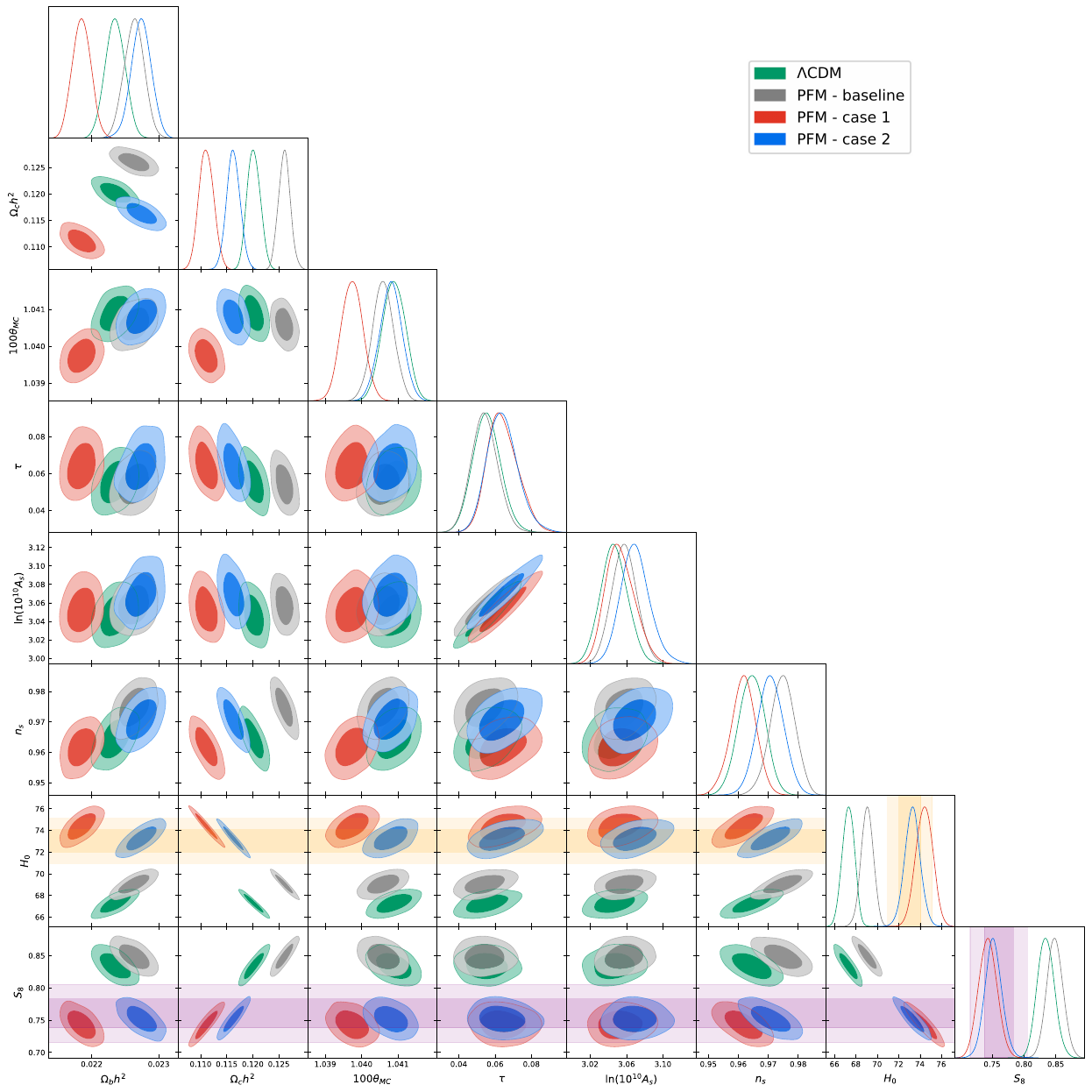}
     \caption{Posterior distributions for the standard model parameters in the PFM with noncanonical sound speeds. For all PFM cases, the background model parameters are set to $n=6$, $\log(10^{10} \Omega_0)=-3.95$, and $a_t=3.1\times 10^{-4}$, and there is no anisotropic shear ($A_\sigma=0$). In red we show case 1 for which there is a constant sound speed set to $c_\phi^2=0$. In blue we show case 2 for which the sound speed is dynamical, given by Eq.~(\ref{eq: dyn c_phi}), and $a_{c_\phi}=a_t$. Posteriors for the $\Lambda$CDM and baseline PFM are shown in green and gray, respectively, for comparison. The darker inner (lighter outer) regions correspond to $1\sigma$($2\sigma$) confidence intervals. The SH0ES Collaboration measurement of $H_0=73.04\pm1.04$ km/s/Mpc and the KiDS-1000 weak lensing survey measurement of $S_8=0.759^{+0.024}_{-0.021}$ are shown in the orange and purple bands, respectively \cite{Riess:2021jrx,KiDS:2020suj}. Distributions are generated with the P18 dataset.
     }
     \label{fig: vary sound speed P18}
 \end{figure*}
 \begin{figure}[t]
     \centering
     \includegraphics[width=\linewidth]{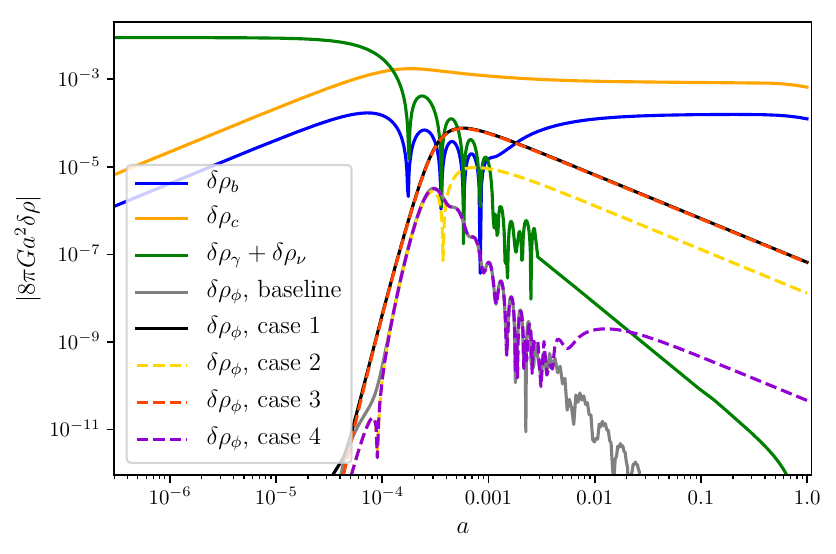}
     \caption{The evolution of the density perturbation of all relevant components as a function of scale factor for the $k=0.1$ Mpc$^{-1}$ wave mode. These curves are generated from a shear-less model with $n=6$, $\log(10^{10}\Omega_0)= -3.95$, and $a_t=3.1 \times 10^{-4}$ with the standard model parameters set to their $\Lambda$CDM best-fit values. The density perturbation of our phenomenological fluid $\delta\rho_\phi$ is shown in case 1 (black) and the baseline model (gray) for comparison. Setting $c_\phi^2=0$ results in the unhindered growth of density fluctuations at later times as seen by the black curve.}
     \label{fig: dgrho}
 \end{figure}

We have shown that a phenomenological fluid model can solve the Hubble tension in the most minimal microphysics scenario with a canonical sound speed of $c_\phi^2=1$ and no anisotropic shear. Before including anisotropic shear in our microphysics scenario, we investigate how changing only the effective sound speed of the fluid alters cosmological parameter estimation, while holding $A_\sigma=0$. The background model parameters are fixed to $n=6$, $\log(10^{10}\Omega_0)=-3.95$, and $a_t=3.1\times 10^{-4}$ for ease in comparison to the baseline case. We then vary the sound speed from its baseline value of $c_\phi^2=1$, considering the four alternative cases, each with no added anisotropic shear, outlined in Table \ref{tab: summary of models}. We consider two cases for comparison. The first is $\Lambda$CDM, used as a control. The second is the baseline model, used for direct comparison of the altered microphysics. 

The results of the MCMC analysis, consisting of constraints on cosmological parameters for cases 1-4 are presented in Table \ref{tab: vary sound speed P18}. We show the posterior distributions for the relevant parameters in these models in Fig.~\ref{fig: vary sound speed P18}. We restrict our dataset to only include \textit{Planck} 2018 data to focus on the CMB inference of $H_0$ within this phenomenological fluid. We present results for extended datasets in Appendix \ref{sec: extended results}. 

We can see from Table \ref{tab: vary sound speed P18} that setting $c_\phi^2=0$ in case 1 gives values of $H_0$ and $S_8$ that are in better agreement with local measurements than the baseline model. For case 1, the Hubble tension is reduced even further from the baseline model to $< 1 \sigma$, with a best-fit value of $H_0=73.97$ km/s/Mpc, compared with the SH0ES Collaboration measurement of $H_0=73.2\pm1.3$ km/s/Mpc, shown by the orange bands in Fig.~\ref{fig: vary sound speed P18} \cite{Riess:2020fzl}. Case 1 also offers a complete resolution to the $S_8$ tension, with a best-fit value of $S_8=0.749$, compared to the measurement from the KiDS-1000 weak lensing survey of $S_8=0.759^{+0.024}_{-0.021}$, shown by the purple bands in Fig.~\ref{fig: vary sound speed P18} \cite{KiDS:2020suj}. While the cosmological parameter estimation may be favorable in case 1, the fit to the data is significantly degraded, with a $\Delta\chi^2_\text{min}=312.38$ compared to the $\Lambda$CDM model. 

To better understand the effect of setting $c_\phi^2=0$ in case 1, we take a closer look at the fluid perturbations equations of motion. Whenever $c_\phi^2=0$, the pressure perturbation simplifies to 
\begin{equation}
\label{eq: delta p}
    \delta p_\phi = -3\mathcal{H}(\rho_\phi+p_\phi)\left[ w_\phi- \frac{w_\phi'}{3 \mathcal{H} (1+w_\phi)}\right]\theta_\phi/k^2 .
\end{equation}
Using this to simplify Eq.~(\ref{eq: theta prime}) we can see that
\begin{equation}
    \theta_\phi'=-\mathcal{H}\theta_\phi,
\end{equation}
meaning that with adiabatic initial conditions, where $\theta_\text{init}=0$, the velocity divergence of our fluid is always zero. With $\theta_\phi=0$, we can see from Eq.~(\ref{eq: delta p}) that $\delta p_\phi=0$, meaning that this fluid clusters. By the same logic, Eq.~(\ref{eq: delta prime}) simplifies to
\begin{equation}
    \delta_\phi' = -(1+w_\phi) \frac{h'}{2} + 3\mathcal{H} w_\phi \delta_\phi, 
\end{equation}
where we see that when $c_\phi^2=0$, there is nothing to damp the growth of density perturbations of our phenomenological fluid. This can be seen in Fig.~\ref{fig: dgrho}, where we show the evolution of the density perturbations of all relevant components as a function of scale factor. Compared to the baseline model shown in gray, the density perturbation of the EDE fluid in case 1, shown in black, is non-negligible at late times, dominating over the radiation components at late times, and over the baryonic contribution for a brief period when the background fluid density first spikes. 

This addition of a clustering fluid component deepens the gravitational potentials and decreases power over the first acoustic peak. The cold dark matter (CDM) and baryon densities in this model must be lowered to account for this additional clustering component as seen in Fig.~\ref{fig: vary sound speed P18}. The change in $\Omega_c h^2$ is also why we see a slightly higher $H_0$ in case 1 than in the baseline model; a lower CDM density shifts the acoustic peaks towards low $\ell$, so to maintain the correct angular scales in the CMB anisotropy pattern, $H_0$ must be raised even more, as seen in Table \ref{tab: vary sound speed P18}. It is important to note that the decrease we see in the $S_8$ parameter in case 1 is mostly driven by the change in the matter density since $S_8=\sigma_8 \sqrt{\Omega_m/0.3}$, where $\sigma_8$ gives the amplitude of matter fluctuations. With a lower matter density, the structure growth parameter $S_8$ is decreased while keeping the actual amplitude of fluctuations $\sigma_8$ effectively fixed. Case 1 gives $\sigma_8=0.8284^{+0.0062}_{-0.0072}$, in good agreement with the baseline constraint of $\sigma_8=0.8305\pm0.0060$.

This description of the $c_\phi^2=0$ model offers insight on the dynamical sound speed model as well. From the constraints presented in Table \ref{tab: vary sound speed P18}, we can see that the baseline case with $c_\phi^2=1$ and case 4 are virtually indistinguishable. In case 4, the sound speed speed does not begin the transition from $c_\phi^2=1$ to $c_\phi^2=0$ until after the equation of state has transitioned to $w_\phi=1$. With virtually no change in the fit to the data from the baseline case, this suggests that once the background fluid density begins to redshift away, the sound speed has little impact. Alternatively, in case 3, $c_\phi^2=0$ well before the fluid becomes dynamical and relevant. As such, the parameter constraints and fit to the data are similar to those for case 1, where $c_\phi^2=0$ always. Case 2 lies in the middle with the transition in the sound speed and equation of state happening simultaneously. In this case, we see the fit to the data begins to degrade and the best-fit parameters shift towards their case 1 values. Looking at the evolution of the density perturbations in each of these cases makes these parameter constraints clear. In Fig.~\ref{fig: dgrho}, we see that $\delta\rho_\phi$ in cases 2 and 3 both dominate over the contribution from baryons and radiation for a brief period, resulting in the same changes to the gravitational potentials that cause the poor fit to the data in case 1. However, $\delta\rho_\phi$ in case 4 only begins to grow once the background density of the fluid is negligible so we do not see the same domination at late times. 

These constraints tell a simple story: the earlier that $c_\phi^2=0$, the longer the fluid clusters and the growth of density perturbations are left unchecked, resulting in a worse fit to the data. After the fluid starts redshifting away, the low background density keeps the sound speed from leaving too strong an imprint. Hence, dynamical $c_\phi(a)$ models with $a_{c_\phi} > a_t$ are effectively the same as the baseline model, and models with $a_{c_\phi}<a_t$ are effectively the same as the $c_\phi^2=0$ model. As the sound speed can only take values of $0\leq c_\phi^2 \leq 1$, setting $c_\phi^2=0$ represents the maximally different case of those we consider here. 

Putting all of these constraints together, it seems that altering the sound speed of EDE from its canonical value of $c_\phi^2=1$ without jointly altering the background dynamics of the fluid, as suggested in Refs.~\cite{Lin:2019qug,Poulin:2018cxd}, is not preferred by \textit{Planck} 2018 data, despite providing preferable constraints on the Hubble constant and structure growth parameter.

\subsection{Shear model I}

 \begin{figure*}[t]
     \centering
     \includegraphics[width=0.9\textwidth]{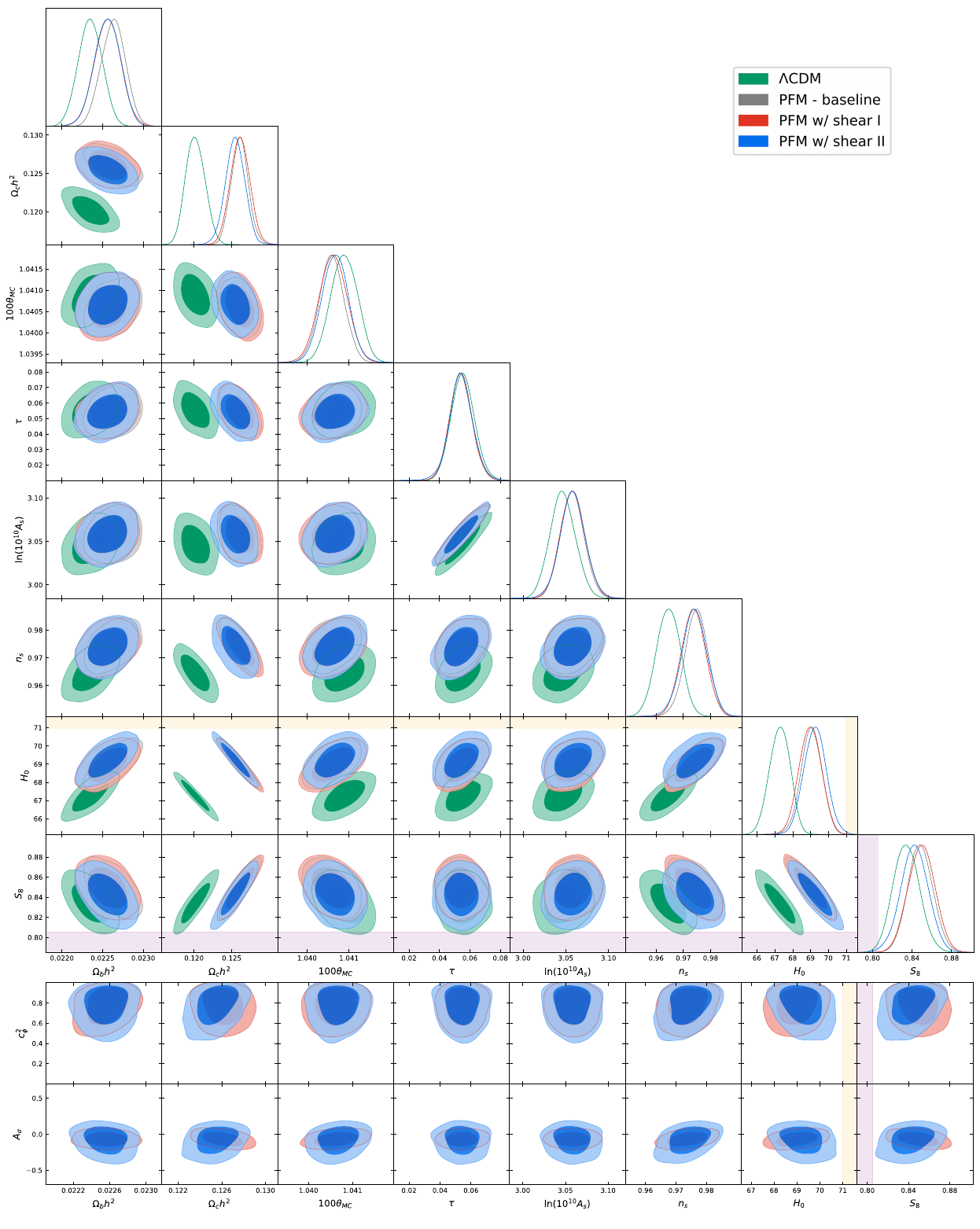}
     \caption{
     Top: Posterior distributions of the standard model parameters for the PFM with shear model I (red) and shear model II (blue) with $n=6$, $\log(10^{10} \Omega_0)=-3.95$, and $a_t=3.1\times 10^{-4}$. Posteriors for the $\Lambda$CDM model (green) and the baseline PFM (gray) are shown for comparison. 
     Bottom: Posterior distributions of the standard model parameters vs the microphysics parameters for the PFM with shear model I (red) and shear model II (blue) with $n=6$, $\log(10^{10} \Omega_0)=-3.95$, and $a_t=3.1\times 10^{-4}$.
     The darker inner (lighter outer) regions correspond to $1\sigma$($2\sigma$) confidence intervals. The SH0ES Collaboration measurement of $H_0=73.04\pm1.04$ km/s/Mpc and the KiDS-1000 weak lensing survey measurement of $S_8=0.759^{+0.024}_{-0.021}$ are shown in the orange and purple bands, respectively \cite{Riess:2021jrx,KiDS:2020suj}. Distributions are generated with the P18 dataset. Both shear models converged to a negligible amount of added shear, with $c_\phi^2$ very nearly equal to unity, making them virtually indistinguishable from the baseline model.
     } 
     \label{fig: shear models P18}
 \end{figure*}
 \begin{figure}[t]
     \centering
     \includegraphics[width=\linewidth]{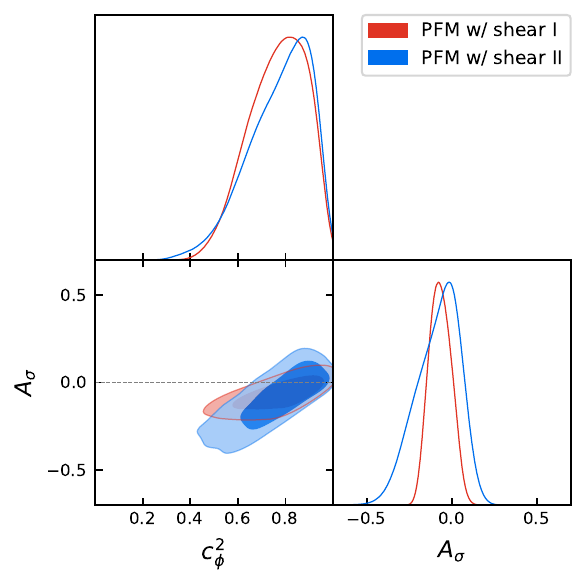}
     \caption{Posterior distributions for the microphysics parameters $c_\phi^2$ and $A_\sigma$ in the PFM model with shear model I (red) and shear model II (blue) with $n=6$, $\log(10^{10} \Omega_0)=-3.95$, and $a_t=3.1 \times 10^{-4}$. The darker inner (lighter outer) regions correspond to $1\sigma$($2\sigma$) confidence intervals. In both cases, the posterior distributions favor $A_\sigma=0$, with a slight preference for negative $A_\sigma$, particularly in shear model II. Distributions are generated with the P18 dataset. } 
     \label{fig: cphi Asig P18}
 \end{figure}
 \begin{figure}[t]
     \centering
     \includegraphics[width=\linewidth]{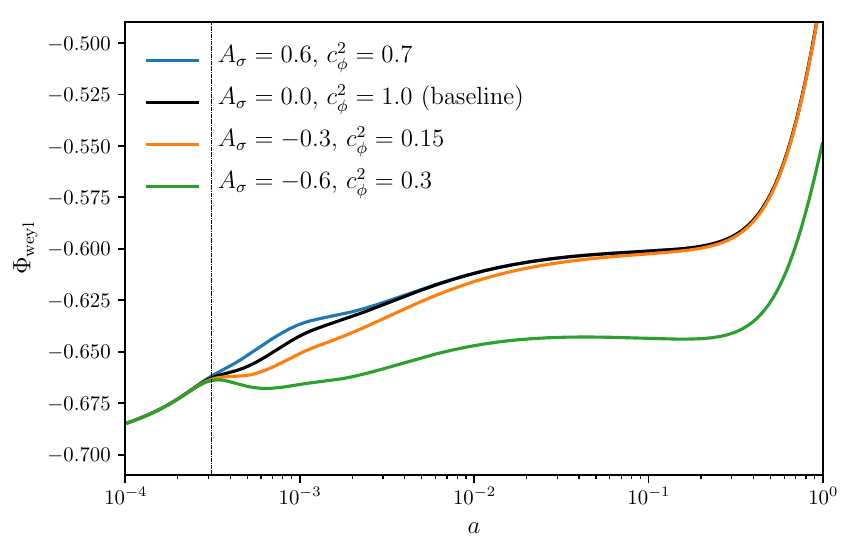}
     \caption{Evolution of the Weyl gravitational potential as a function of scale factor for the $k=7\times 10^{-6}$ Mpc$^{-1}$ wave mode in shear model I. The black curve shows the baseline model with no added shear. The blue (orange) curves show that a case with positive (negative) $A_\sigma$ makes the potential wells shallower (deeper) for the brief period of time that the background density of the EDE fluid is relevant. The green curve shows that for a very anisotropic fluid ($A_\sigma<-0.6$), the Weyl potential diverges from its baseline evolution. The black-dashed line shows the transition scale factor $a_t$, at which the fluid becomes dynamical. }
     \label{fig: weyl cases}
 \end{figure}
 \begin{figure*}[t]
     \centering
     \includegraphics[width=0.9\textwidth]{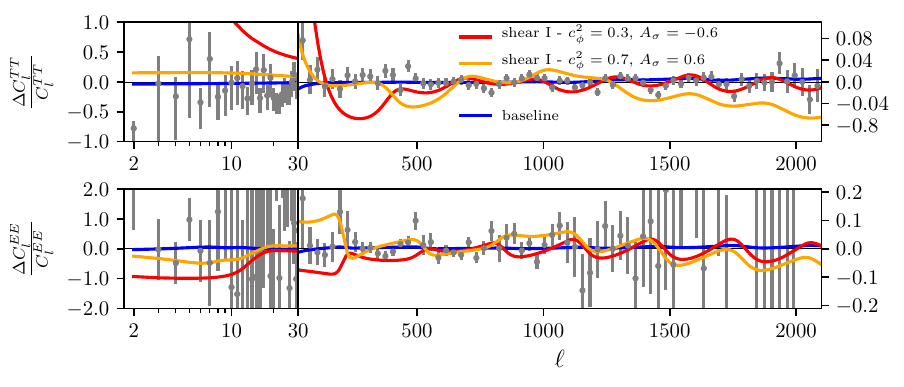}
     \caption{ Temperature and polarization power spectrum residuals between the best-fit $\Lambda$CDM model and the best-fit fluid model with shear model I with positive (orange) and negative (red) $A_\sigma$. The best-fit baseline model is shown in blue for comparison. Residuals from \textit{Planck} 2018 data are shown in gray. Left (right) vertical axis scaling is for multipoles less (greater) than $\ell=30$. }
     \label{fig: residuals shear I}
 \end{figure*}

We have shown that only changing the sound speed of our phenomenological EDE fluid cannot improve the solution to the Hubble tension. We now move on to including anisotropic shear in our model, starting with shear model I, given by Eq.~(\ref{eq: shear model 1}). We hold the background model fixed at $n=6$, $\log(10^{10}\Omega_0)=-3.95$, and $a_t=3.1\times 10^{-4}$ to facilitate direct comparison to the baseline case, and vary $c_\perp^2$ and $c_\parallel^2$ alongside our six $\Lambda$CDM parameters to derive constraints on $c_\phi^2$ and $A_\sigma$. Parameter constraints for this shear model are presented in Table \ref{tab: shear I P18}. Posterior distributions for relevant parameters in this model, along with shear model II, are presented in Fig.~\ref{fig: shear models P18}-\ref{fig: cphi Asig P18}, with $\Lambda$CDM and baseline posteriors included for comparison. 

\begin{table}[t]
\begin{tabular}{| c | c |}
\hline \hline
 Parameter & PFM w/ shear I \\ \hline \hline
 $100 \omega_b$      & $2.256(2.225)\pm 0.016$ \\ 
 $\omega_c$          & $0.1261(0.1262)\pm0.0013$ \\ 
 $100 \theta_s$      & $1.04062(1.04043)\pm0.00033$ \\
 $\tau$              & $0.0541(0.0531)\pm0.0073$ \\ 
 $\ln (10^{10} A_s)$ & $3.058(3.056)\pm0.014$ \\ 
 $n_s$               & $0.9737(0.9727)\pm0.0044$ \\
 $c_\phi^2$          & $0.770(0.848)^{+0.150}_{-0.097}$ \\ 
 $A_\sigma$          & $-0.068(-0.031)^{+0.063}_{-0.070}$ \\ 
 \hline
 $H_0$ [km/s/Mpc]    & $68.97(68.83)\pm0.62$ \\ 
 $S_8$               & $0.850(0.850)\pm0.013$ \\
 \hline
 Total $\chi^2_\text{min}$ & 1013.31 \\
 $\Delta \chi^2_\text{min}$ & -0.78 \\
 \hline \hline 
\end{tabular}
\caption{\label{tab: shear I P18} The mean (best-fit) $\pm 1\sigma$ error of the cosmological parameters for phenomenological fluid model with shear model I. The background model parameters are held fixed at $n=6$, $\log(10^{10} \Omega_0)=-3.95$, and $a_t=3.1\times10^{-4}$. Constraints are derived from the P18 dataset and $\Delta\chi^2_\text{min}$ is calculated with respect to the best-fit $\Lambda$CDM model presented in Table \ref{tab: background constraints}.}
\end{table}

As we can see from Table \ref{tab: shear I P18}, the best-fit microphysics parameters for this shear model make it virtually indistinguishable from the baseline case with $c_\phi^2=0.848$ and $A_\sigma=-0.031$. To get a clear picture as to why this model of anisotropic shear is so disfavored by data, it is useful to take a closer look at how the addition of this shear changes the CMB angular power spectrum via its interactions with the other perturbative quantities of the EDE fluid. 

The magnitude and sign of the EDE shear is controlled via $A_\sigma$. The total shear in the cosmic fluid $\pi_{tot}$, is given by the sum of all nonzero shear components $\pi_{tot}= \Sigma_i (1+w_i)\sigma_i$. Besides our EDE component, the only other shear contributions come from radiation for which $\sigma_\gamma,\sigma_\nu<0$. This means that for a model with $A_\sigma=0$, $\pi_{tot}<0$. When $(1+w_\phi)\sigma_\phi>0$, due to a positive $A_\sigma$, the EDE shear adds destructively to the total shear in the system, lowering the magnitude of $\pi_{tot}$. Conversely, when $(1+w_\phi)\sigma_\phi<0$, due to a negative $A_\sigma$, the EDE shear enhances the total shear in the system, increasing the magnitude of $\pi_{tot}$. These changes to the total shear contribution compared to the baseline model have a significant impact on the Weyl potential $\Phi$, given by 
\begin{equation}
    \Phi = -\frac{8\pi G}{2k^2}a^2 \sum_i \rho_i \left[\delta_i+3\mathcal{H}(1+w_i)\theta_i/k^2 +\frac{3}{2}(1+w_i)\sigma_i\right], 
\end{equation}
where $i$ sums over all components of the total energy density. Hence, when we increase or decrease the magnitude of the total shear contribution, the gravitational potentials get deeper or shallower, respectively. 

In addition to the inherent impact of adding a new component to the total shear, $\sigma_\phi$ influences the evolution of the velocity perturbation of the EDE via Eq.~(\ref{eq: theta prime}), which in turn alters the evolution of the density perturbation. At large scales, we can analytically solve for the scaling behavior of these perturbations which we parametrize via an effective equation of state such that $\delta_\phi\propto a^{-3(1+w_\delta)}$ and $\theta_\phi \propto a^{-3(1+w_\theta)}$. During matter domination we find that 
\begin{equation}
    w_\delta = -\frac{1}{4}\left(5-2 A_\sigma+\sqrt{9+4A_\sigma(A_\sigma-1)-8c_\phi^2}\right),
\end{equation}
\begin{equation}
     w_\theta = -\frac{1}{12}\left(17-6 A_\sigma+3\sqrt{9+4A_\sigma(A_\sigma-1)-8c_\phi^2}\right), 
\end{equation}
which tell us that as you increase $A_\sigma$, both $\delta_\phi$ and $\theta_\phi$ decay more rapidly, regardless of the background equation of state of the fluid. 

For very positive $A_\sigma$, the density and velocity perturbations decay quicker than the baseline $A_\sigma=0$ case, causing the overall magnitude of the Weyl potential to be decreased, and vice versa for negative $A_\sigma$ cases. The overall changes to the Weyl potential due to the inherent impact of $\sigma_\phi$ and the subsequent impact on $\delta_\phi$ and $\theta_\phi$ can be seen in Fig.~\ref{fig: weyl cases} where we plot the evolution of the Weyl potential at low-$k$ for different manifestations of shear model I. We can see that for very positive $A_\sigma$, shown in blue, the gravitational potentials are shallower than the baseline model, shown in black, for the brief period following the transition in the background equation of state when the EDE fluid has a non-negligible background abundance. Due to the rapid scaling of the perturbations and the background behavior of the fluid density, this suppression of the Weyl potential is short lived, and the gravitational potentials return to their baseline trajectory at late times. When $A_\sigma<0$, we see the opposite behavior. The Weyl potential is deepened, and because $\delta_\phi$ and $\theta_\phi$ do not decay as quickly as they do in the baseline model, their effect lasts longer. When $A_\sigma\lesssim -0.6$, the fluid perturbations dominate over the standard model components, causing the Weyl potential to diverge from its baseline trajectory as seen through the green curve in Fig.~\ref{fig: weyl cases}. 

For fixed $c_\phi^2$ and $A_\sigma$, this behavior at large scales leads to large changes at $\ell\lesssim 1000$ in the CMB angular power spectrum as seen in Fig.~\ref{fig: residuals shear I} where we plot the residuals between the best-fit $\Lambda$CDM model and shear model I, for choices of positive and negative $A_\sigma$. We can see that for $A_\sigma=-0.6$, shown in red, the divergence of the Weyl potential suppresses power over the first acoustic peak and enhances power at $\ell\lesssim 100$, when compared to the baseline model. These large changes to the power spectrum for non-negligible amounts of shear explain the best-fit parameters for shear model I presented in Table \ref{tab: shear I P18}. \textit{Planck} 2018 data constrains the amount of shear allowed in this model to be very close to zero with $A_\sigma=-0.068^{+0.063}_{-0.070}$, suggesting that the large-scale influence of this equation-of-state formulation of shear on gravitational potentials is too large an obstacle to overcome. Putting this all together, if EDE has some anisotropic shear component, the data suggests it should not be introduced in the form of Eq.~(\ref{eq: shear model 1}) due to the large-scale influence of the shear on the evolution of gravitational potentials. 

\subsection{Shear model II}
\label{sec: shear II results}
 \begin{figure*}[t]
     \centering
     \includegraphics[width=0.9\textwidth]{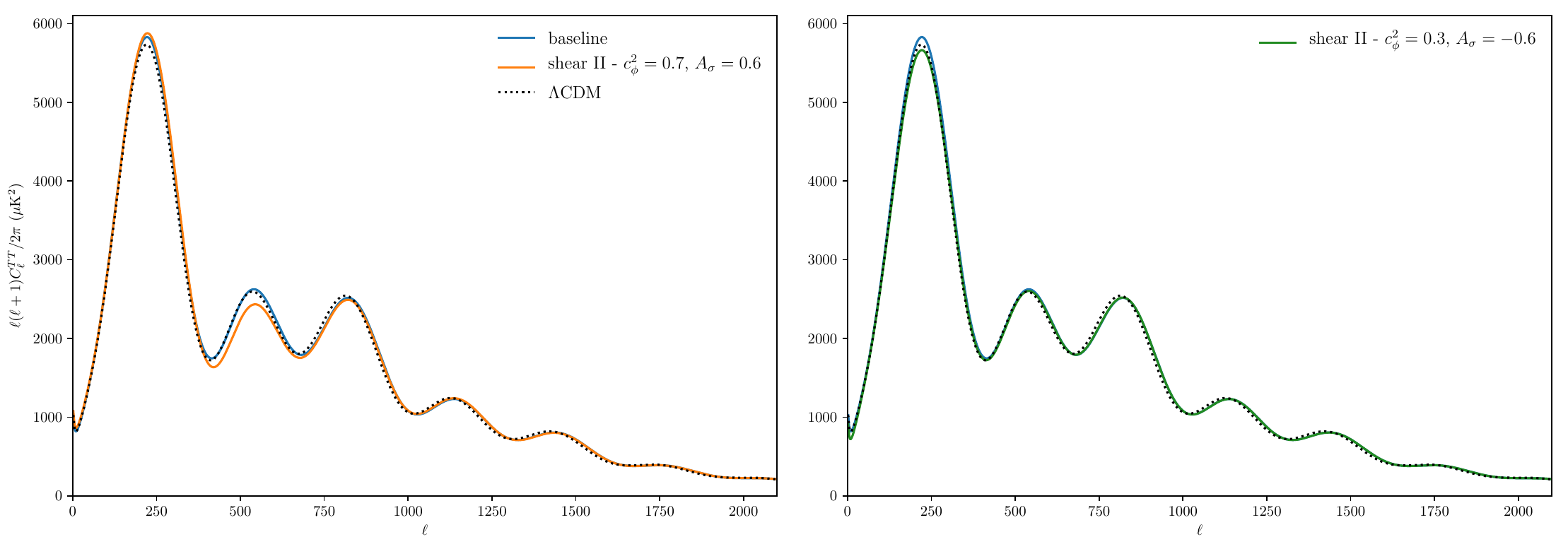}
     \caption{Temperature power spectrum for shear model II. All curves are generated with the standard model parameters set to their best-fit $\Lambda$CDM values from Table \ref{tab: background constraints}. For the PFM models we set $n=6$, $a_t=3.1\times 10^{-4}$, and $\log(10^{10}\Omega_0)=-3.95$. The blue curves shows the baseline model with $c_\phi^2=1$, and $A_\sigma=0$, the orange curve shows shear model II with a fixed positive $A_\sigma$, and the green curve shows shear model II with a fixed negative $A_\sigma$. }
     \label{fig: shear II TT spectrum}
 \end{figure*}
 \begin{figure*}[t]
     \centering
     \includegraphics[width=0.9\textwidth]{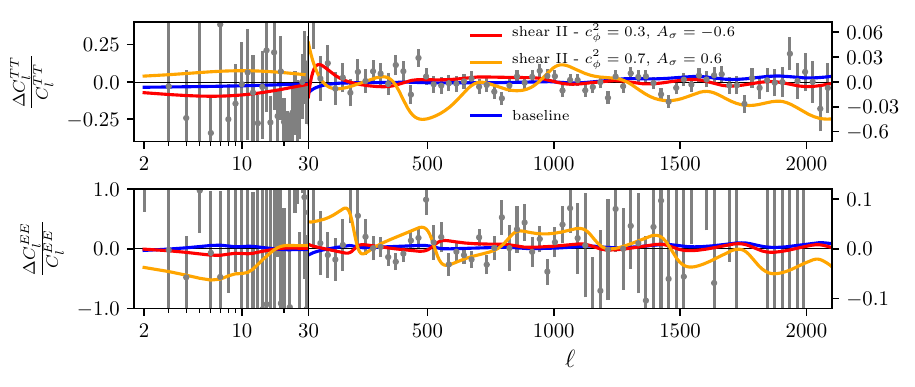}
     \caption{Temperature and polarization power spectrum residuals between the best-fit $\Lambda$CDM model and the best-fit fluid model with shear model II with positive (orange) and negative (red) $A_\sigma$. The best-fit baseline model is shown in blue for comparison. Residuals from \textit{Planck} 2018 data are shown in gray. Left (right) vertical axis scaling is for multipoles less (greater) than $\ell=30$.}     \label{fig: residuals shear II}
 \end{figure*}
\begin{table*}[t]
\begin{tabular}{| c | c | c | c | c |}
\hline \hline
 Parameter &  PFM w/ shear II & shear II - $A_\sigma=0.6$ & shear II - $A_\sigma=-0.6$ & shear II - $A_\sigma=-0.2$ \\ \hline \hline
 $100 \omega_b$      & $2.256(2.257)\pm0.016$ & $2.069(2.070)\pm0.012$ & $2.275(2.269)\pm0.016$ & $2.250(2.250)\pm0.015$\\ 
 $\omega_c$          & $0.1255(0.1257)\pm0.0013$ & $0.1260(0.1259)\pm0.0011$ & $0.1245(0.1247)\pm0.0012$ & $0.1245(0.1248)\pm0.0012$\\ 
 $100 \theta_s$      & $1.04067(1.04059)\pm0.00031$ & $1.04007(1.04009)\pm0.00032$ & $1.04026(1.04032)\pm0.00030$ & $1.04067(1.04065)\pm 0.00030$\\
 $\tau$              & $0.0541(0.0569)\pm0.0078$ & $0.0391(0.0419)^{+0.0089}_{-0.0072}$ & $0.0546(0.0528)\pm0.0075$ & $0.0550(0.0562)\pm0.0074$\\ 
 $\ln (10^{10} A_s)$ & $3.058(3.060)\pm0.015$ & $3.035(3.039)^{+0.017}_{-0.014}$ & $3.063(3.062)\pm0.014$ & $3.060(3.063)\pm0.014$\\ 
 $n_s$               & $0.9740(0.9772)\pm0.0047$ & $0.9591(0.9603)\pm0.0038$ & $0.9618(0.9602)\pm0.0041$ & $0.9723(0.9715)\pm0.0042$\\
 $c_\phi^2$          & $0.778(0.951)^{+0.170}_{-0.087}$ & 0.7 (fixed) & 0.3 (fixed) & 0.55 (fixed)\\ 
 $A_\sigma$          & $-0.08(0.08)^{+0.15}_{-0.10}$ & 0.6 (fixed) & -0.6 (fixed) & -0.2 (fixed) \\ 
 \hline
 $H_0$ [km/s/Mpc]    & $69.24(69.11)\pm0.62$ & $67.07(67.13)\pm0.53$ & $69.64(69.52)\pm0.61$ & $69.56(69.41)\pm0.60$\\ 
 $S_8$               & $0.843(0.847)\pm0.014$ & $0.863(0.864)\pm0.012$ & $0.830(0.8311)\pm0.013$ & $0.833(0.838)\pm0.013$\\
 \hline
 Total $\chi^2_\text{min}$ & 1013.48 & 1355.69 & 1031.73 & 1016.50\\
 $\Delta \chi^2_\text{min}$ & -0.61 & 341.60 & 17.64 & 2.41\\
 \hline \hline 
\end{tabular}
\caption{\label{tab: shear II P18} The mean (best-fit) $\pm 1\sigma$ error of the cosmological parameters for phenomenological fluid model with shear model shear model II with $n=6$, $\log(10^{10} \Omega_0)=-3.95$, $a_t=3.1\times10^{-4}$, and different choices of $c_\phi^2$ and $A_\sigma$. Constraints are derived from the P18 dataset and $\Delta\chi^2_\text{min}$ is calculated with respect to the best-fit $\Lambda$CDM model presented in Table \ref{tab: background constraints}. The best-fit values for the $A_\sigma=0.6$ and $A_\sigma=-0.6$ cases were used to generate the orange and red curves in Fig.~\ref{fig: residuals shear I}, respectively.}
\end{table*}

Our second shear model is defined via a physically motivated \cite{Hu:1998kj}, gauge-invariant equation of motion given by Eq.~(\ref{eq: shear model 2}). Similarly to shear model I, the best-fit parameters for shear model II with free $c_\phi^2$ and $A_\sigma$, given in Table \ref{tab: shear II P18}, are statistically indistinguishable from the best-fit baseline model, given in Table \ref{tab: baseline P18}. However, as can be seen by the blue curve in Fig.~\ref{fig: cphi Asig P18}, the $1\sigma$ constraints on the microphysics parameters are much looser in shear model II than they are in shear model I. Specifically, a degeneracy between $c_\phi^2$ and $A_\sigma$ exists allowing a non-negligible amount of negative shear coupled with a lower effective sound speed. 

These constraints are made clearer by looking at the effect of positive and negative $A_\sigma$ values on the CMB angular power spectrum. Figure \ref{fig: shear II TT spectrum} shows the temperature power spectrum for cases with very positive $A_\sigma=0.6$ (orange) and very negative $A_\sigma=-0.6$ (green), with the baseline case and the best-fit $\Lambda$CDM model shown in blue, and black, respectively, for comparison. We show parameter constraints for shear model II with the same positive and negative $A_\sigma$ in Table \ref{tab: shear II P18}. 

In the baseline case, which mimics standard EDE, power is enhanced over the first acoustic peak and all peaks are shifted towards large scales. These changes to the power spectrum result in the parameter shifts seen in the best-fit baseline model shown in Table \ref{tab: baseline P18}, most notably, increased $\omega_c$ and $H_0$ values. On top of the changes to the power spectrum we see in the baseline case, when $A_\sigma$ is positive, we see an added suppression of power over the second acoustic peak. This suppression requires a higher value of $\omega_b$ as seen in Table \ref{tab: shear II P18}, restoring the heights of the first and second acoustic peaks into agreement with \textit{Planck} 2018 data, in conjunction with changes to the CDM density. The key difference between this model and the baseline case is that the requirement of a higher baryon density also shifts the acoustic peaks towards larger scales, relinquishing the need for a higher value of the Hubble constant, resulting in a best-fit value of $H_0=67.13$ km/s/Mpc, virtually unchanged from the best-fit $\Lambda$CDM value. Overall, these parameter changes lead to a poor fit to the data seen through the residuals between this positive $A_\sigma$ case and the best-fit $\Lambda$CDM model shown by the orange curve in Fig.~\ref{fig: residuals shear II}. 

Turning to a negative shear case, we get a different story. From Fig.~\ref{fig: cphi Asig P18} we know that \textit{Planck} 2018 data allows for a non-negligible, but not large, amount of negative shear. For explanatory purposes we focus on an extremal case with $A_\sigma=-0.6$ and $c_\phi^2=0.3$ to show the full effect of a negative shear in this model. From Fig.~\ref{fig: shear II TT spectrum}, we can see that when we add in this negative shear, leaving all $\Lambda$CDM parameters unchanged from their best-fit values, the enhancement over the first acoustic peak that comes in the baseline model is avoided, giving a temperature power spectrum whose main difference from the $\Lambda$CDM model is a shift in all acoustic peaks towards small scales. As can be seen in Table \ref{tab: shear II P18}, this leads to an even higher Hubble constant than the baseline case with $H_0=69.64\pm0.61$ km/s/Mpc. Similarly to the baseline case, the CDM density must be increased from its $\Lambda$CDM value, but in this model the added shear counteracts the deepening of the gravitational potentials caused by the increase in the matter density, leaving the $S_8$ value in statistical agreement with $\Lambda$CDM giving $S_8=0.830\pm0.013$. So, in addition to strengthening the solution to the $H_0$ tension, this model does not exacerbate the $S_8$ tension like standard EDE models. As can be seen in Fig.~\ref{fig: residuals shear II}, the best-fit model with $A_\sigma=-0.6$ results in a slightly poorer fit to \textit{Planck} 2018 data than the baseline case with a $\Delta\chi^2_\text{min}=17.64$, as is to be expected for this extremal case. 

\begin{figure*}[t]
     \centering
     \includegraphics[width=0.9\textwidth]{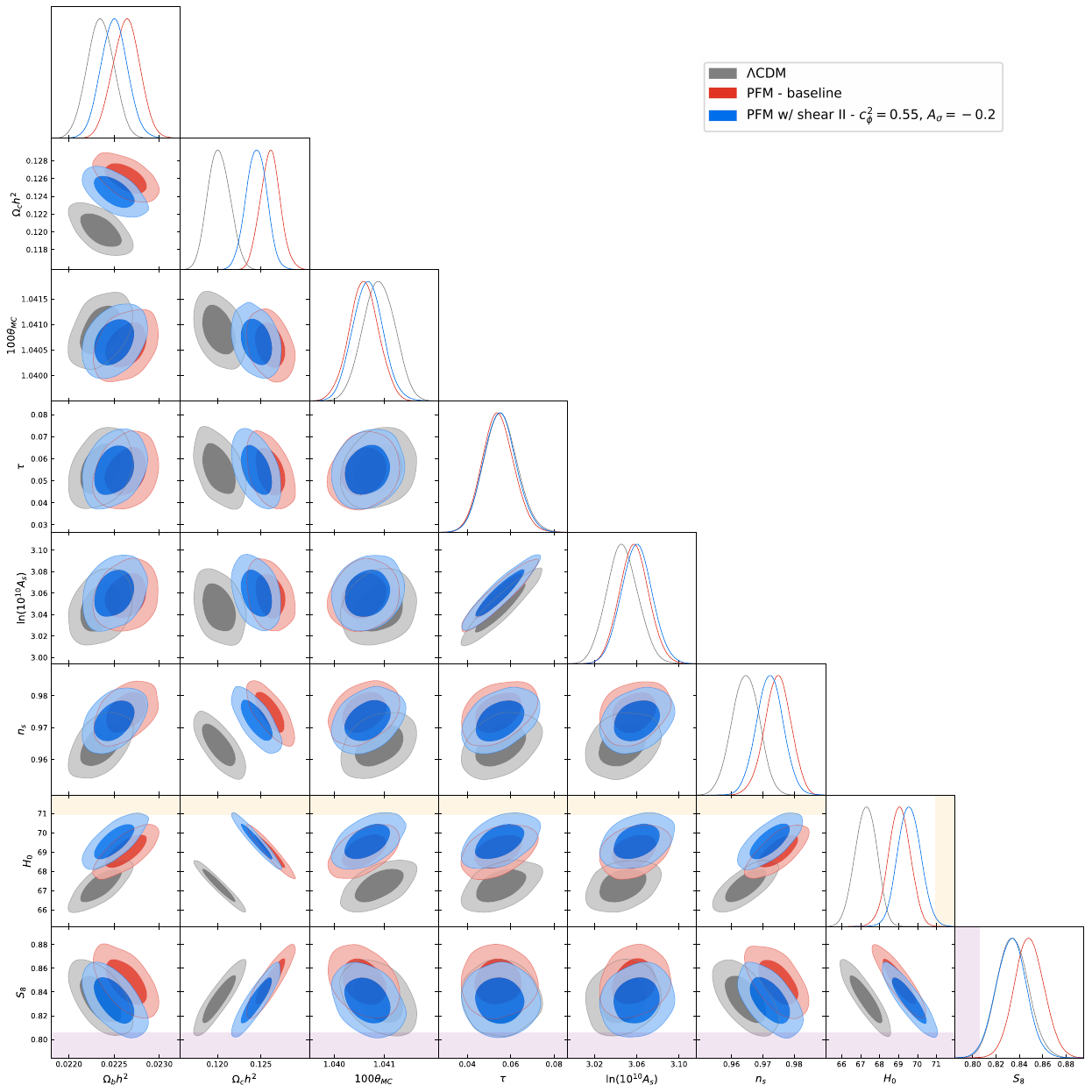}
     \caption{ Posterior distributions for the PFM model with shear model II (blue) with $n=6$, $\log(10^{10} \Omega_0)=-3.95$, $a_t=3.1\times 10^{-4}$, $c_\phi^2=0.55$, and $A_\sigma=-0.2$. The $\Lambda$CDM (gray) and baseline models (red) are shown for comparison. 
     The darker inner (lighter outer) regions correspond to $1\sigma$($2\sigma$) confidence intervals. The SH0ES collaboration measurement of $H_0=73.04\pm1.04$ km/s/Mpc and the KiDS-1000 weak lensing survey measurement of $S_8=0.759^{+0.024}_{-0.021}$ are shown in the orange and purple bands, respectively \cite{Riess:2021jrx,KiDS:2020suj}. Distributions are generated with the P18 dataset. A small, but non-negligible amount of negative shear added to a generic EDE model can simultaneously soften the $H_0$ and $S_8$ tensions in comparison to standard EDE \cite{Smith:2019ihp}.} 
     \label{fig: good model P18}
 \end{figure*}

For a more reasonable choice of $c_\phi^2=0.55$ and $A_\sigma=-0.2$, which lies within the $2\sigma$ contours in Fig.~\ref{fig: cphi Asig P18}, we see the same shifts in $H_0$ and $S_8$ as we do in the extremal case, shown in Fig.~\ref{fig: good model P18}, but with a comparable fit to the data as $\Lambda$CDM. From these results we see that the addition of a negative shear that evolves according to the equation of motion given in Eq.~(\ref{eq: shear model 2}) strengthens EDE as a solution to the Hubble tension. In this case, the $S_8$ tension is not exacerbated by the inclusion of our new component while preserving the solution to the Hubble tension. 

This can be seen more clearly in comparison with local measurements of $S_8$ and $H_0$. Comparing our $S_8$ constraints to the combined DES-Y3 constraint of $S_8=0.776\pm0.017$ \cite{DES:2021wwk}, we find that our baseline EDE model with a best-fit $S_8=0.849$ gives a $\chi^2_\text{DES}=18.44$, whereas shear model II with $A_\sigma=-0.2$ gives $\chi^2_\text{DES}=13.30$ with its best-fit value of $S_8=0.838$. Compared with the $\Lambda$CDM model best-fit value of $S_8=0.829$ and $\chi^2_\text{DES}=9.72$, our negative shear model offers a significant improvement over standard EDE. 

Similarly, as a resolution to the Hubble tension, we see a slightly better fit with our negative shear model to the SH0ES Collaboration measurement of $H_0=73.04\pm1.04$ km/s/Mpc \cite{Riess:2021jrx} which for $\Lambda$CDM leads to a $\chi^2_\text{SH0ES}=28.89$. The baseline EDE model softens this tension with a best-fit $H_0=69.11$ km/s/Mpc leading to a $\chi^2_\text{SH0ES}=14.28$. Shear model II with $A_\sigma=-0.2$ improves on this slightly with $H_0=69.41$ km/s/Mpc giving $\chi^2_\text{SH0ES}=12.18$, a slight, but statistically irrelevant, improvement to the resolution given by the baseline EDE model.

We have also considered the case where the background parameters, $r_\phi$, $a_t$, and $n$, are allowed to vary. We fix the microphysics parameters to $c_\phi^2=0.55$ and $A_\sigma=-0.2$ to explore the broader impact of this negative shear model. This method has the advantage of providing constraints that consider the full range of background evolution possible with this microphysics scenario. However, as with the standard EDE fluid model discussed in Sec.~\ref{sec: resolution to the H0 tension}, one must use the full dataset, in particular a late-universe prior on $H_0$, in order to find preference for a nonzero density of EDE. While the constraints on $S_8$ and $H_0$ in this extended case are similar to the case with a fixed background evolution run on only \textit{Planck} data, the $\Lambda$CDM constraint on $S_8$ from the full dataset, shown in Table \ref{tab: background constraints}, is lower than the $\Lambda$CDM constraint from \textit{Planck} data alone. This results in a weaker softening of the $S_8$ tension when the background parameters are sampled over, but one that is still statistically relevant. For a more extended discussion of this scenario see Appendix \ref{sec: extended results}. Any solution to the cosmological tensions will preferably exist in \textit{Planck} data alone. For this reason, we fix the background evolution in our main analysis. With a fixed background evolution we focus on the effects of EDE microphysics on cosmological parameter constraints from \textit{Planck} data specifically.

This anisotropic microphysics scenario may be a sign of nonscalar field EDE. However, this region of parameter space is indistinguishable from standard scalar field EDE when considering \textit{Planck} data alone, so we must look to future experiments to provide meaningful constraints on the microphysics of EDE. 

\subsection{Future constraints}
\begin{figure}[t]
     \centering
     \includegraphics[width=\linewidth]{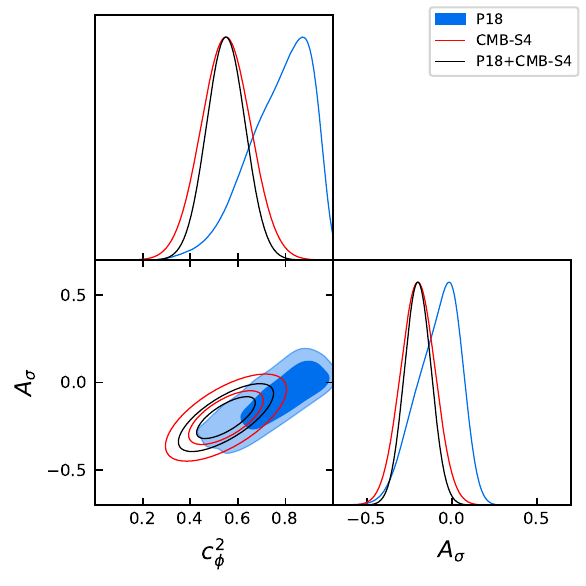}
     \caption{Forecasts for $c_\phi^2$ and $A_\sigma$ for the CMB-S4 experiment (red) and its combination with \textit{Planck} (black). The blue contour shows current constraints using \textit{Planck} 2018 data alone (same as Fig.~\ref{fig: cphi Asig P18}). With or without the inclusion of the P18 dataset, CMB-S4 will be able to distinguish a case with a small, but non-negligible amount of shear from the baseline, shear-less case.} 
     \label{fig: cmb-s4 cphi2 Asig}
 \end{figure}

\begin{table}[b]
\begin{tabular}{| c | c | c |}
\hline \hline
 Parameter           & Fiducial &  CMB-S4 \\ \hline \hline
 $100 \omega_b$      & 2.250    & $\pm 0.006$ \\ 
 $\omega_c$          & 0.1248   & $\pm0.0025$ \\ 
 $H_0$               & 69.41    & $\pm0.93$ \\
 $10^9 A_s$          & 2.140    & $\pm 0.014$ \\ 
 $n_s$               & 0.9715   & $\pm0.0039$  \\ 
 $\tau$              & 0.0562   & $\pm 0.0027$ \\
 $n$                 & 6        & $\pm 0.09$ \\
 $a_t \times 10^4$   & 3.1      & $\pm 0.6$ \\
 $\log(10^{10} \Omega_0)$ & -3.95 & $\pm 1.08$ \\
 $c_\phi^2$          & 0.55 & $\pm 0.104$ \\ 
 $A_\sigma$          & -0.2 & $\pm 0.103$ \\ 
 \hline \hline 
\end{tabular}
\caption{\label{tab: cmb-s4 constraints } Forecasted $1\sigma$ parameter constraints for the PFM model with shear model II assuming a CMB-S4 experiment.}
\end{table}
We forecast constraints on this model using a Fisher information matrix formalism assuming a CMB-S4 experiment that covers 40\% of the sky, following the prescription laid out in Ref.~\cite{CMB-S4:2016ple}. We model our Gaussian noise according to
\begin{equation}
    N_\ell^{\alpha\alpha} = \Delta^2 \exp\left(\ell(\ell+1)\frac{\theta_\text{FWHM}^2}{8 \ln{2}}\right)
\end{equation}
where $\alpha\in\{T,E\}$, $\Delta$ is the white noise level in $\mu$K-arcmin, and $\theta_\text{FWHM}$ is the beam width. We consider a telescope beam with $\theta_\text{FWHM}=1'$ and a white noise level of $\Delta_T=1$ $\mu$K' for temperature and $\Delta_E=\sqrt{2} \Delta_T$ for polarization. We compute the covariance matrix as
\begin{multline}
    \mathbb{C}_\ell(C_\ell^{\alpha\beta},C_\ell^{\gamma\delta}) = \frac{1}{(2\ell+1)f_\text{sky}} [ (C_\ell^{\alpha\gamma}+N_\ell^{\alpha\gamma})(C_\ell^{\beta\delta}+N_\ell^{\beta\delta}) \\  +(C_\ell^{\alpha\delta}+N_\ell^{\alpha\delta})(C_\ell^{\beta\gamma}+N_\ell^{\beta\gamma}) ],
\end{multline}
where $\alpha,\beta,\gamma,\delta \in \{T,E\}$, and $f_\text{sky}$ is the fractional sky coverage of the CMB-S4 experiment considered. Finally the Fisher matrix is calculated using 
\begin{equation}
    F_{ij}=\sum_\ell \frac{\partial{C_\ell^\top}}{\partial{\theta_i}}\mathbb{C}_\ell^{-1}\frac{\partial{C_\ell}}{\partial{\theta_j}}, 
\end{equation}
where $\theta_i$ runs over the six $\Lambda$CDM parameters, as well as our five model parameters $n$, $a_t$, $\log(10^{10}\Omega_0)$, $c_\perp^2$, and $c_\parallel^2$, making $F_{ij}$ an 11x11 matrix. Table \ref{tab: cmb-s4 constraints } gives the fiducial model used in our Fisher analysis, as well as the forecasted $1\sigma$ constraints on all parameters. Figure \ref{fig: cmb-s4 cphi2 Asig}, shows the forecasted posterior distributions for $c_\phi^2$ and $A_\sigma$ assuming the fiducial model. From Fig.~\ref{fig: cmb-s4 cphi2 Asig} we can see that CMB-S4 should be able to distinguish the case with $c_\phi^2=0.55$ and $A_\sigma=-0.2$ from the standard EDE model where $c_\phi^2=1$ and $A_\sigma=0$. 
It is important to note that a Fisher matrix formalism assumes Gaussian errors for all parameters. As suggested by the constraints on $c_\phi^2$ and $A_\sigma$ presented in Fig.~\ref{fig: cphi Asig P18} and Fig.~\ref{fig: cmb-s4 cphi2 Asig}, the underlying probability distribution for individual parameters in our model may not be Gaussian. Hence the $1\sigma$ error predictions from our Fisher forecast given in Table \ref{tab: cmb-s4 constraints } should not be thought of as restrictive constraints. Nevertheless, they serve as useful references on the ability of CMB-S4 to constrain new physics. Assuming the fiducial model and errors presented in Table \ref{tab: cmb-s4 constraints }, CMB-S4 may be able to distinguish the underlying microphysics at the $4\sigma$ level. If future constraints do favor $A_\sigma\neq 0$, this would be evidence of a richer microphysics sector than that implied by scalar field EDE.

In short, EDE with an anisotropic shear in the form of Eq.~(\ref{eq: shear model 2}), with $c_\phi^2\sim0.55$ and $A_\sigma\sim -0.2$, can reduce the Hubble tension to $<3\sigma$, while not exacerbating the $S_8$ tension like standard EDE models. The region of microphysics parameter space that accomplishes this solution is indistinguishable from a shear-less case with current data, but the CMB-S4 experiment will increase precision, allowing us to concretely assess the viability of altering the microphysics of EDE. 

\section{Discussion}
\label{sec: conclusions}

Early dark energy has emerged as one of the most promising classes of solutions to the Hubble tension, however the microphysics of the canonical scalar fields used in these models preclude fully satisfactory solutions mainly by exacerbating the $S_8$ tension even further. In this paper we investigate the ability of noncanonical microphysics to strengthen EDE as a solution to the Hubble tension. We describe EDE as a phenomenological fluid component whose background evolution mimics standard EDE, and alter the perturbative dynamics of the fluid by allowing the effective sound speed of the fluid to differ from its canonical value of $c_\phi^2=1$, and by introducing an anisotropic shear perturbation via an equation of state formalism (shear model I) or an equation of motion (shear model II). In total this phenomenological model constitutes a five parameter extension to $\Lambda$CDM, with three parameters to describe the background evolution $n$, $a_t$, and $\log(10^{10}\Omega_0)$, and two parameters to describe the microphysics $c_\phi^2$, and $A_\sigma$. 

We find that for models with no added anisotropic shear, the $H_0$ and $S_8$ tensions can be jointly ameliorated by making the phenomenological fluid cluster through setting $c_\phi^2=0$ before the transition in the background equation of state. However, only altering the sound speed leads to a significantly worse fit to \textit{Planck} 2018 data, making this an unfavorable solution to the tensions. This poor fit comes in response to the deepening of gravitational potentials caused by the addition of a new clustering component. Models that transition from a nonclustering to clustering fluid, thereby
limiting the time that the clustering can effect the gravitational potentials, suffer the same problem, unless the transition in the sound speed happens well after the fluid density begins to redshift away with $w_\phi=1$.

Furthermore, we find that the inclusion of anisotropic shear can help or hinder EDE as a solution to the Hubble tension, depending on the way it is introduced. For shear model I, defined by the gauge-invariant equation of state given in Eq.~(\ref{eq: shear model 1}), the addition of a new shear component to the total stress energy of the system leads to significant changes to the evolution of the density and velocity perturbations of the fluid, and of the evolution of the Weyl potential at large scales. These large-scale changes to the perturbative evolution lead to significant alterations to the CMB angular power spectrum at $\ell<1000$, which in turn constrain the amount of shear allowed in this model to be negligible.

Alternatively, when anisotropic shear is introduced via the equation of motion given in Eq.~(\ref{eq: shear model 2}), we find significantly different results. For a non-negligible region of parameter space, the inclusion of this shear in a generic EDE model not only slightly improves upon the resolution to the Hubble tension provided by EDE, but simultaneously softens the $S_8$ tension with $H_0=69.56\pm0.60$ km/s/Mpc and $S_8=0.833\pm0.013$, when compared to the standard EDE case which gives $H_0=69.11\pm0.58$ km/s/Mpc and $S_8=0.849\pm0.013$. This favorable region of parameter space loosely given by $c_\phi^2>0.6$ and $A_\sigma<-0.2$, is indistinguishable from the standard EDE model using \textit{Planck} 2018 data alone. Using a Fisher information matrix analysis, we found that future observations from CMB-S4 will be able to distinguish between these different microphysical scenarios. 

A clear preference for a non-negligible amount of EDE shear would imply that if EDE is at play, it need not be the result of a canonical scalar field. 
Rather, strongly anisotropic microphysics may be indicative of a novel component that is isotropic at the background level, but breaks isotropy perturbatively. Examples range from free-streaming neutrinos to more speculative models such as a cosmic lattice or coherent vector fields. Our approach has been to study the impact of the equation of state, sound speed, and anisotropic shear more generally.

While our focus has been on the scalar sector, it is reasonable to expect that any microphysical model that gives rise to scalar anisotropic stress will also contribute vector and tensor stress. The latter is of great interest, for the potential to affect a B-mode polarization signal of primordial gravitational waves. A wide range of behavior may be expected, considering free-streaming neutrinos \cite{Weinberg:2003ur}, topological defects \cite{Turok:1997gj}, and coherent vector fields \cite{Bielefeld:2015daa}. We leave this subject for later investigation.

Future probes of LSS and the CMB will be essential to verifying if anisotropic EDE was present in the early universe, and will offer further clues into the microphysics of EDE. 

\begin{acknowledgments}
We thank Jose Luis Bernal, Vivian Poulin, and Tristan Smith for useful comments. This work is supported in part by U.S. Department of Energy Award No. DE-SC0010386. Computing resources provided in part by the National Energy Research Scientific Computing Center (NERSC), a U.S. Department of Energy Office of Science User Facility located at Lawrence Berkeley National Laboratory.
\end{acknowledgments}

\appendix

\section{Shear Model Derivation}
\label{sec: shear model details}
In this Appendix we give extended derivations of the shear models presented in Sec.~\ref{sec: shear models}.
\subsection{Shear model I}
Following \cite{Ma:1995ey}, the velocity divergence of a single uncoupled fluid, like our phenomenological EDE fluid, can be written most generally as 
\begin{equation}
\label{eq: general theta eom}
    \theta' = -\mathcal{H}(1-3w)\theta-\frac{w'}{1+w}\theta+\frac{k^2}{1+w}\frac{\delta p}{\rho} - k^2\sigma,
\end{equation}
where the pressure is given by Eq.~(\ref{eq: pressure pert}). From this equation we can see that the pressure perturbation $\delta p$ and anisotropic shear $\sigma$ act as positive and negative source terms respectively. We specifically design our shear equation of state in shear model I to counteract the growth of the pressure source term in Eq.~(\ref{eq: general theta eom}). We define the shear to be
\begin{equation}
\label{eq: GI stress 1}
    (\rho+p)\sigma = c_\sigma^2 (\delta p -c_t^2 \delta \rho - 3\mathcal{H}(c_t^2-c_a^2)(\rho+p)\theta/k^2),
\end{equation}
where $c_\sigma^2$ and $c_t^2$ are new parameters. For $c_\sigma^2=1$ and $c_t^2=0$, the pressure perturbation source term in Eq.~(\ref{eq: general theta eom}) is completely canceled, however we choose to leave them as free parameters for completeness. The second and third terms in Eq.~(\ref{eq: GI stress 1}) are included to keep the stress equation of state gauge invariant. Plugging Eq.~(\ref{eq: pressure pert}) into Eq.~(\ref{eq: GI stress 1}) we find 
\begin{equation}
    (\rho+p)\sigma = c_\sigma^2 (c_\phi^2-c_t^2) (\delta\rho+3\mathcal{H}(\rho+p)\theta/k^2). 
\end{equation}
By setting $A_\sigma = c_\sigma^2 (c_\phi^2-c_t^2)$, we recover Eq.~(\ref{eq: shear model 1}) which we use to define shear model I. 

If we directly substitute shear model I into the equation of motion for the velocity perturbation we find
\begin{equation}
    \label{eq: theta prime with shear I}
    \theta'=-\mathcal{H}\left[1-3(c_\phi^2-A_\sigma)\right]\theta+\frac{k^2}{1+w}(c_\phi^2-A_\sigma)\delta,
\end{equation}
where it becomes clear that for $c_\phi^2=A_\sigma$, we get complete cancellation of the source term for the velocity perturbation making $\theta=0$ at all times with adiabatic initial conditions. 

\subsection{Shear model II}

Our second model of shear is derived directly from the density and velocity perturbations whose evolution equations are given by Eqs.~(\ref{eq: delta prime}) and (\ref{eq: theta prime}). We start by differentiating Eq.~(\ref{eq: theta prime}) with respect to conformal time to get a second order equation giving
\begin{multline}
    \theta''=-\mathcal{H}'(1-3c_\phi^2)\theta -\mathcal{H}(1-3c_\phi^2)\theta'+\frac{c_\phi^2}{1+w}k^2\delta'\\ -3\mathcal{H}(w-c_a^2)\frac{c_\phi^2}{1+w}k^2\delta -k^2\sigma'
\end{multline}
where we have assumed a time-varying equation of state and used Eq.~(\ref{eq: ca2 def}) to simplify. Using Eqs.~(\ref{eq: delta prime}) and (\ref{eq: theta prime}) we can write this second order equation as a function exclusively of $\theta$ and $\sigma$,
\begin{multline}
    \theta'' =  -k^2c_\phi^2\frac{h'}{2} [-\mathcal{H}(1-3c_\phi^2)-3\mathcal{H}(c_\phi^2-c_a^2)]\theta'\\  -[k^2c_\phi^2+\mathcal{H}'(1-3c_\phi^2)+3\mathcal{H}^2(c_\phi^2-c_a^2)]\theta\\ - k^2 [\sigma'+3\mathcal{H}(c_\phi^2-c_a^2)\sigma]. 
\end{multline}
At small scales, this simplifies to 
\begin{equation}
    \theta'' = -k^2 c_\phi^2(\theta+\frac{h'}{2}) - k^2[\sigma'+3\mathcal{H}(c_\phi^2-c_a^2)\sigma].
\end{equation}
Now we suppose that
\begin{equation}
    \sigma' + 3\mathcal{H}(c_\phi^2-c_a^2)\sigma = B_\sigma (\theta+\alpha k^2)
\end{equation}
where $\alpha = (h'+6\eta')/2k^2$ with $h$ and $\eta$ being the synchronous gauge metric potentials. The right-hand side of this equation is directly taken from the shear terms in the second order differential we derived for the velocity perturbation. At small scales, this implies
\begin{equation}
    \theta'' = -k^2 (c_\phi^2+B_\sigma)\theta - k^2[(c_\phi^2+B_\sigma)\frac{h'}{2}+3B_\sigma \eta']. 
\end{equation}
For a wave travelling in the $\hat z$ direction, $\theta=\partial_zv^z$, which coupled with the above equation implies $c_\parallel^2=c_\phi^2+B_\sigma$. For cohesion between our shear models, we reparametrize and define $B_\sigma=-A_\sigma$ such that $c_\parallel^2=c_\phi^2 - A_\sigma$, just like in shear model I, which gives us our definition of shear model II presented in Eq.~(\ref{eq: shear model 2}).

\section{Extended Results}
\label{sec: extended results}

In this Appendix we present extended MCMC results from our analysis of the phenomenological EDE fluid model with varied microphysics. In Table \ref{tab: shear I cases P18} we give the parameter constraints for the positive and negative $A_\sigma$ cases run on the P18 dataset used to produce the residuals seen in Fig.~\ref{fig: residuals shear I} for shear model I. As explained in Sec.~\ref{sec: results}, the large scale impact of the anisotropic shear in this model causes the poor fits we see in Table \ref{tab: shear I cases P18}, and constrains the amount of shear allowed to be negligible. 
\begin{table*}[t]
\begin{tabular}{| c | c | c |}
\hline \hline
 Parameter &  $A_\sigma=0.6$ & $A_\sigma=-0.6$ \\ \hline \hline
 $100 \omega_b$ & $2.047(2.050)\pm0.011$ & $2.129(2.123)\pm0.012$ \\ 
 $\omega_c$ & $0.1209(0.1210)\pm0.0012$ & $0.1455(0.1457)\pm0.0010$\\ 
 $100 \theta_s$ & $1.03956(1.03951)\pm0.00029$ & $1.03775(1.03773)\pm0.00030$\\
 $\tau$ & $0.0424(0.0429)^{+0.0075}_{-0.0067}$ & $<0.0125(0.0102)$\\ 
 $\ln (10^{10} A_s)$ & $3.031(3.033)\pm0.015$ & $2.9958(2.9931)^{+0.0058}_{-0.0071}$\\ 
 $n_s$ & $0.9512(0.9507)\pm0.0037$ & $0.9548(0.9536)\pm0.0038$\\
 $c_\phi^2$ & 0.7 (fixed) & 0.3 (fixed) \\ 
 $A_\sigma$ & 0.6 (fixed) & -0.6 (fixed) \\ 
 \hline
 $H_0$ [km/s/Mpc] & $68.73(68.67)\pm0.58$ & $60.02(59.91)\pm0.40$\\ 
 $S_8$ & $0.813(0.816)\pm0.013$ & $1.056(1.057)\pm0.012$\\
 \hline
 Total $\chi^2_\text{min}$ & 1472.76 & 1719.88 \\
 $\Delta \chi^2_\text{min}$ & +458.67 & +705.79 \\
 \hline \hline 
\end{tabular}
\caption{\label{tab: shear I cases P18} The mean (best-fit) $\pm 1\sigma$ error of the cosmological parameters for phenomenological fluid model with shear model I, $n=6$, $\log(10^{10} \Omega_0)=-3.95$, $a_t=3.1\times10^{-4}$, and different choices of $c_\phi^2$ and $A_\sigma$, generated from the P18 dataset. The best-fit values were used to generate the orange and red curves in Fig.~\ref{fig: residuals shear I}. }
\end{table*}

\begin{figure*}[t]
     \centering
     \includegraphics[width=0.9\textwidth]{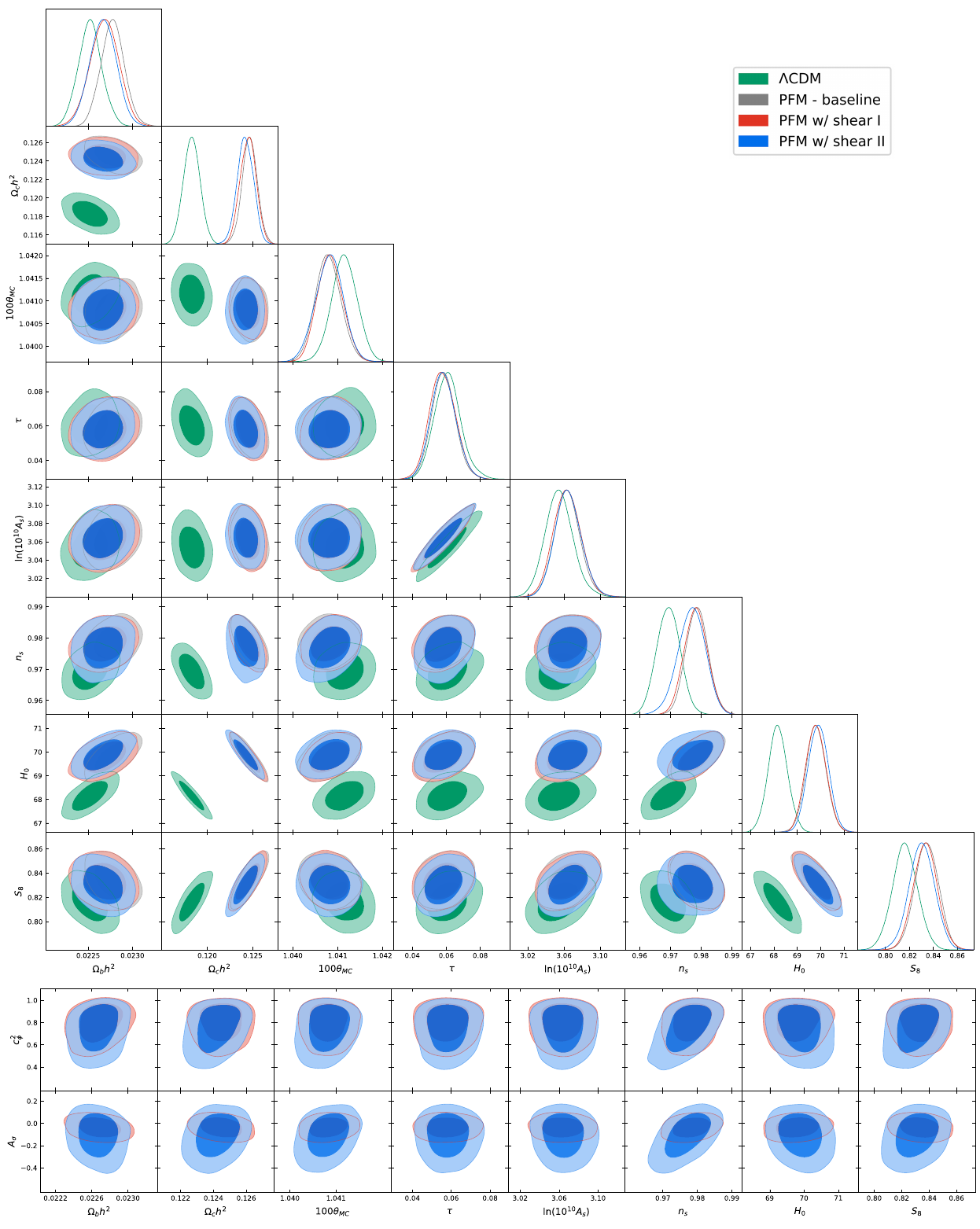}
     \caption{Same as Fig.~\ref{fig: shear models P18} but for the P18+BAO+R19+SN datasets.}
     \label{fig: shear models full}
 \end{figure*}

\begin{figure}[t]
     \centering
     \includegraphics[width=\linewidth]{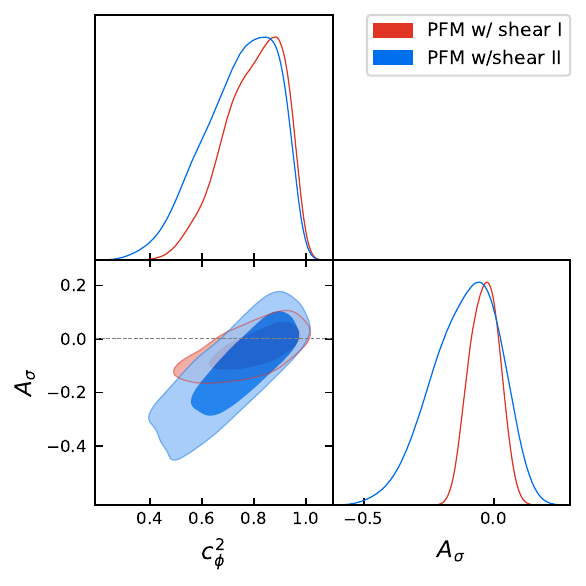}
     \caption{Same as Fig.~\ref{fig: cphi Asig P18} but for the P18+BAO+R19+SN datasets.  
     } 
     \label{fig: cphi Asig full}
 \end{figure}

In Table \ref{tab: shear full}, we show constraints on the cosmological parameters in the baseline PFM, the PFM w/ shear model I and the PFM w/ shear model II run using the P18+BAO+R19+SN datasets. Posterior distributions for the relevant parameters in these models are shown in Figs.~\ref{fig: shear models full} and \ref{fig: cphi Asig full}, with $\Lambda$CDM shown for comparison. Comparing to Tables \ref{tab: shear I P18} and \ref{tab: shear II P18}, we can see that the inclusion of additional datasets does allow for slightly more anisotropic shear with a higher $H_0$, but does not significantly change the results for either model of shear. 
\begin{table*}[t]
\begin{tabular}{| c | c | c | c |}
\hline \hline
 Parameter &  PFM - baseline & PFM w/ shear I & PFM w/ shear II \\ \hline \hline
 $100 \omega_b$ & $2.278(2.280)\pm0.013$ & $2.268(2.273)\pm0.016$ & $2.267(2.272)\pm0.015$ \\ 
 $\omega_c$ & $0.12466(0.12464)\pm0.00083$ & $0.12449(0.12462)\pm0.00087$ & $0.12418(0.12435)\pm0.00088$ \\ 
 $100 \theta_s$ & $1.04079(1.04076)\pm0.00028$ & $1.04083(1.04082)^{+0.00027}_{-0.00030}$ & $1.04081(1.04101)\pm0.00030$ \\
 $\tau$ & $0.0584(0.0582)\pm0.0071$ & $0.0578(0.0598)\pm0.0076$ & $0.0583(0.05711)^{+0.0068}_{-0.0077}$ \\ 
 $\ln (10^{10} A_s)$ & $3.064(3.065)\pm0.014$ & $3.063(3.067)\pm0.015$ & $3.064(3.062)^{+0.014}_{-0.015}$ \\ 
 $n_s$ & $0.9786(0.9775)\pm0.0036$ & $0.9782(0.9783)\pm0.0037$ & $0.9768(0.9786)^{+0.0046}_{-0.0042}$ \\
 $c_\phi^2$ & -  & $0.799(0.787)^{+0.140}_{-0.085}$ & $0.748(0.812)^{+0.180}_{-0.096}$ \\ 
 $A_\sigma$ & -  & $-0.036(-0.058)\pm0.058$ & $-0.11(-0.06)^{+0.15}_{-0.11}$ \\ 
 \hline
 $H_0$ [km/s/Mpc] & $69.80(69.81)\pm0.42$ & $69.79(69.78)\pm0.43$ & $69.89(69.95)\pm0.43$ \\ 
 $S_8$ & $0.8344(0.8342)\pm0.0098$ & $0.833(0.836)\pm0.010$ & $0.830(0.830)\pm0.011$ \\
 \hline
 Total $\chi^2_\text{min}$ & 2063.81 & 2063.59 & 2064.02 \\
 $\Delta \chi^2_\text{min}$ & -9.56 & -9.78 & -9.35 \\
 \hline \hline 
\end{tabular}
\caption{\label{tab: shear full} The mean (best-fit) $\pm 1\sigma$ error of the cosmological parameters in the baseline model, the PFM with shear I, and the PFM with shear II, generated from the P18+BAO+R19+SN datasets. }
\end{table*}

\begin{figure*}[t]
     \centering
     \includegraphics[width=0.9\textwidth]{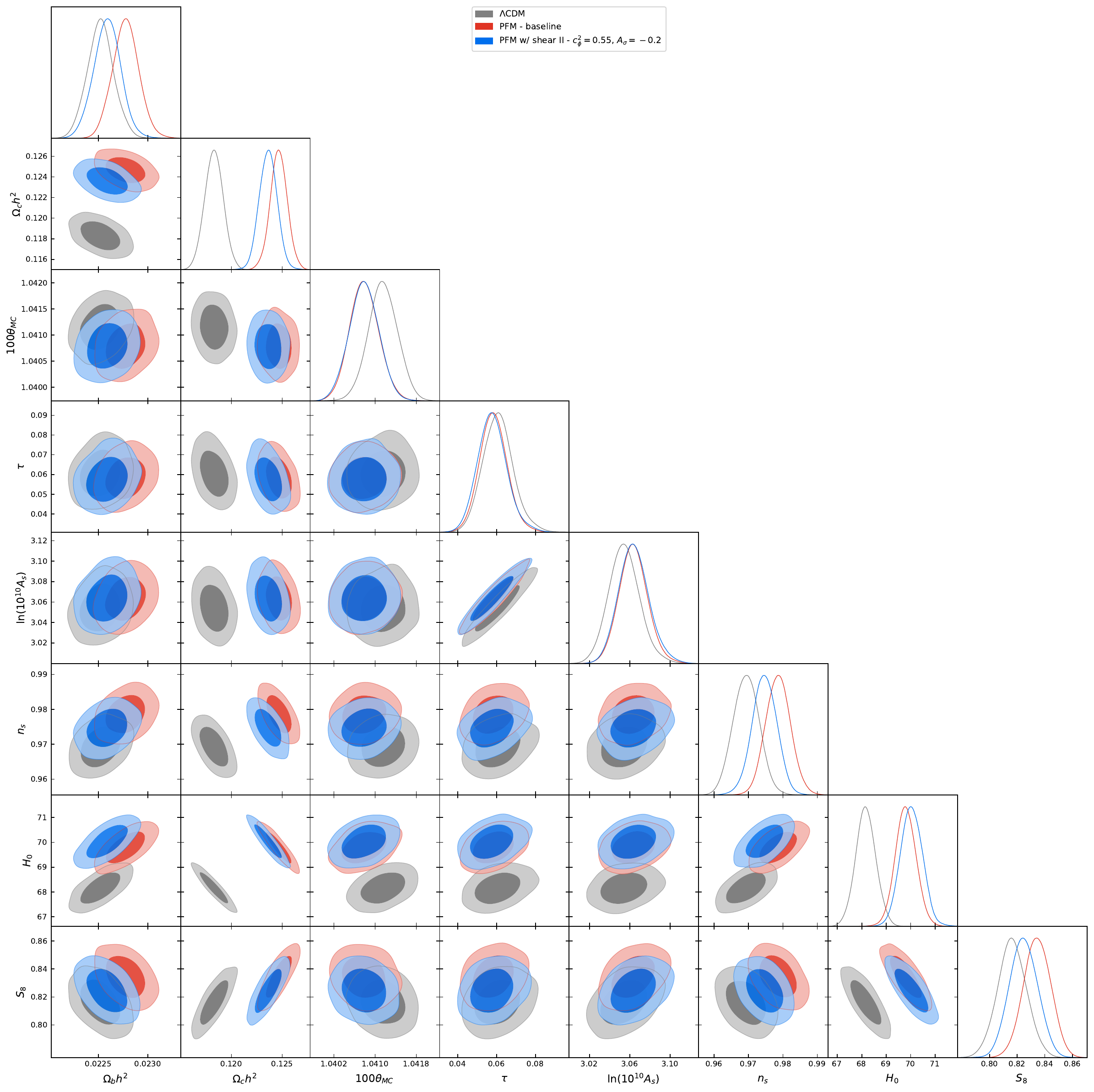}
     \caption{Same as Fig.~\ref{fig: good model P18} but for the P18+BAO+R19+SN datasets. }
     \label{fig: shear II good full}
 \end{figure*}

Next, we give the results of an MCMC analysis, consisting of constraints on cosmological parameters in Table \ref{tab: shear II good full} and posterior distributions for those parameters in Fig.~\ref{fig: shear II good full} , for the phenomenological EDE fluid with shear model II. We show the case of $c_\phi^2=0.55$ and $A_\sigma=-0.2$, discussed in Sec.~\ref{sec: results}, run on the P18+BAO+R19+SN datasets. Similarly to the previous cases, the inclusion of more datasets does not drastically alter the results of the analysis. The main differences are slight upwards and downwards shifts in the posteriors for $H_0$ and $S_8$, respectively when compared to the run with only \textit{Planck} 2018 data. However, in comparison to the best-fit $\Lambda$CDM model run on the same extended dataset, the resolutions to the $H_0$ and $S_8$ tensions are less pronounced. 
\begin{table}[t]
\begin{tabular}{| c | c |}
\hline \hline
 Parameter & PFM w/ shear II - $A_\sigma=-0.2$  \\ \hline \hline
 $100 \omega_b$ & $2.259(2.256)\pm0.014$ \\ 
 $\omega_c$ & $0.12362(0.12379)\pm0.00087$ \\ 
 $100 \theta_s$ & $1.04079(1.04080)\pm0.00029$ \\
 $\tau$ & $0.0579(0.0540)^{+0.0069}_{-0.0078}$ \\ 
 $\ln (10^{10} A_s)$ & $3.064(3.058)^{+0.014}_{-0.016}$ \\ 
 $n_s$ & $0.9746(0.9751)\pm0.0036$ \\
 $c_\phi^2$ & 0.55 (fixed) \\ 
 $A_\sigma$ & -0.2 (fixed) \\ 
 \hline
 $H_0$ [km/s/Mpc] & $70.03(69.94)\pm0.45$ \\ 
 $S_8$ & $0.825(0.824)\pm0.010$ \\
 \hline
 Total $\chi^2_\text{min}$ & 2065.55 \\
 $\Delta \chi^2_\text{min}$ & -7.82 \\
 \hline \hline 
\end{tabular}
\caption{\label{tab: shear II good full} The mean (best-fit) $\pm 1\sigma$ error of the cosmological parameters for phenomenological fluid model with shear model II, $n=6$, $\log(10^{10} \Omega_0)=-3.95$, $a_t=3.1\times10^{-4}$, $c_\phi^2=0.55$, and $A_\sigma=-0.2$, generated from the P18+BAO+R19+SN datasets.}
\end{table}

\begin{figure*}[t]
     \centering
     \includegraphics[height=0.9\textheight]{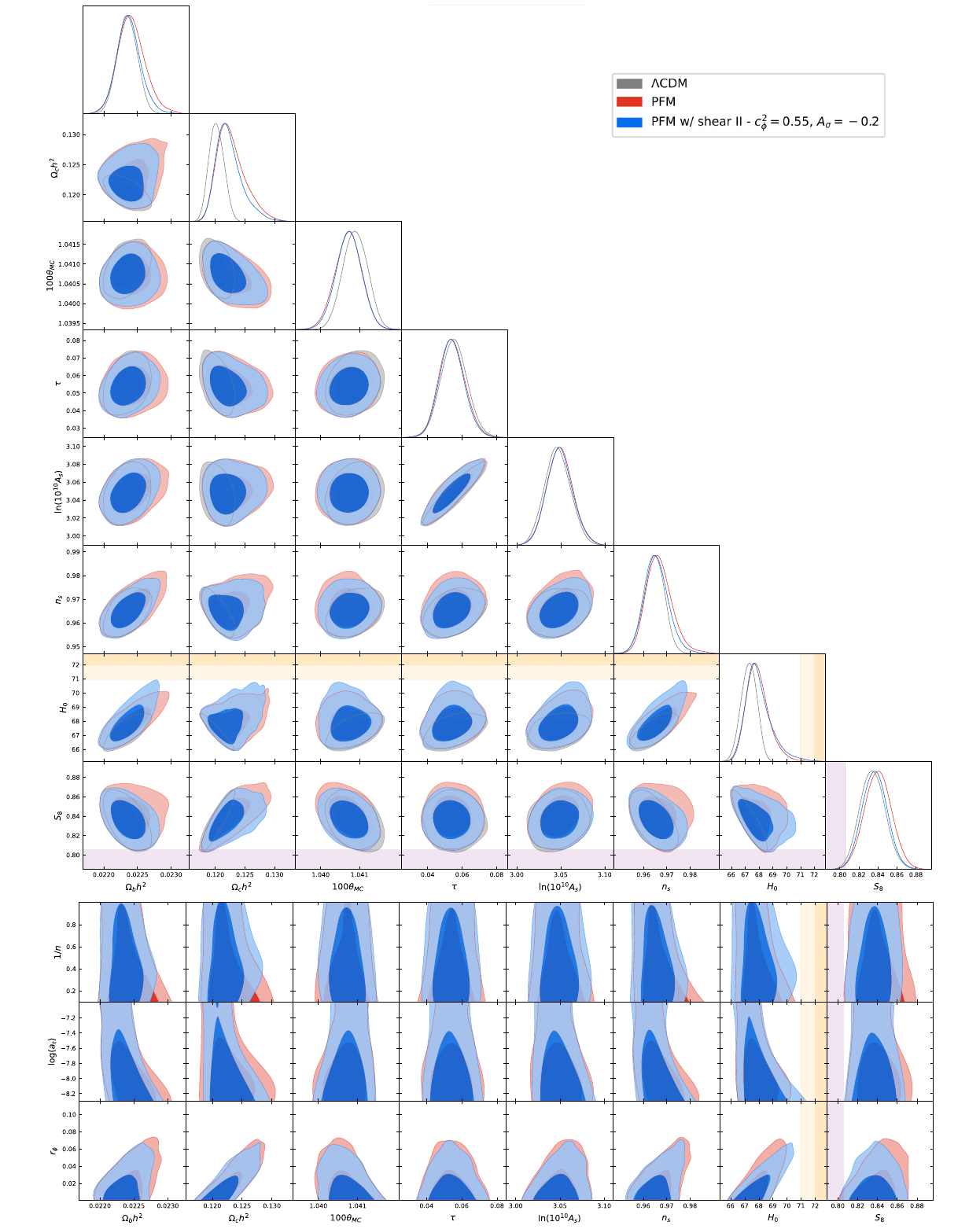}
     \caption{
     Top: Posterior distributions of the standard model parameters for the $\Lambda$CDM model (gray), the PFM with $c_\phi^2=1$ and $A_\sigma=0$ (red), and shear model II with $c_\phi^2=0.55$ and $A_\sigma=-0.2$ (blue). Bottom: Posterior distributions of the standard model parameters vs the background PFM parameters. The darker inner (lighter outer) regions correspond to $1\sigma$($2\sigma$) confidence intervals. The SH0ES Collaboration measurement of $H_0=73.04\pm1.04$ km/s/Mpc and the KiDS-1000 weak lensing survey measurement of $S_8=0.759^{+0.024}_{-0.021}$ are shown in the orange and purple bands, respectively \cite{Riess:2021jrx,KiDS:2020suj}. Distributions are generated with the P18 dataset.
     }
     \label{fig: shearII background P18}
 \end{figure*}

\begin{figure*}[t]
     \centering
     \includegraphics[height=0.9\textheight]{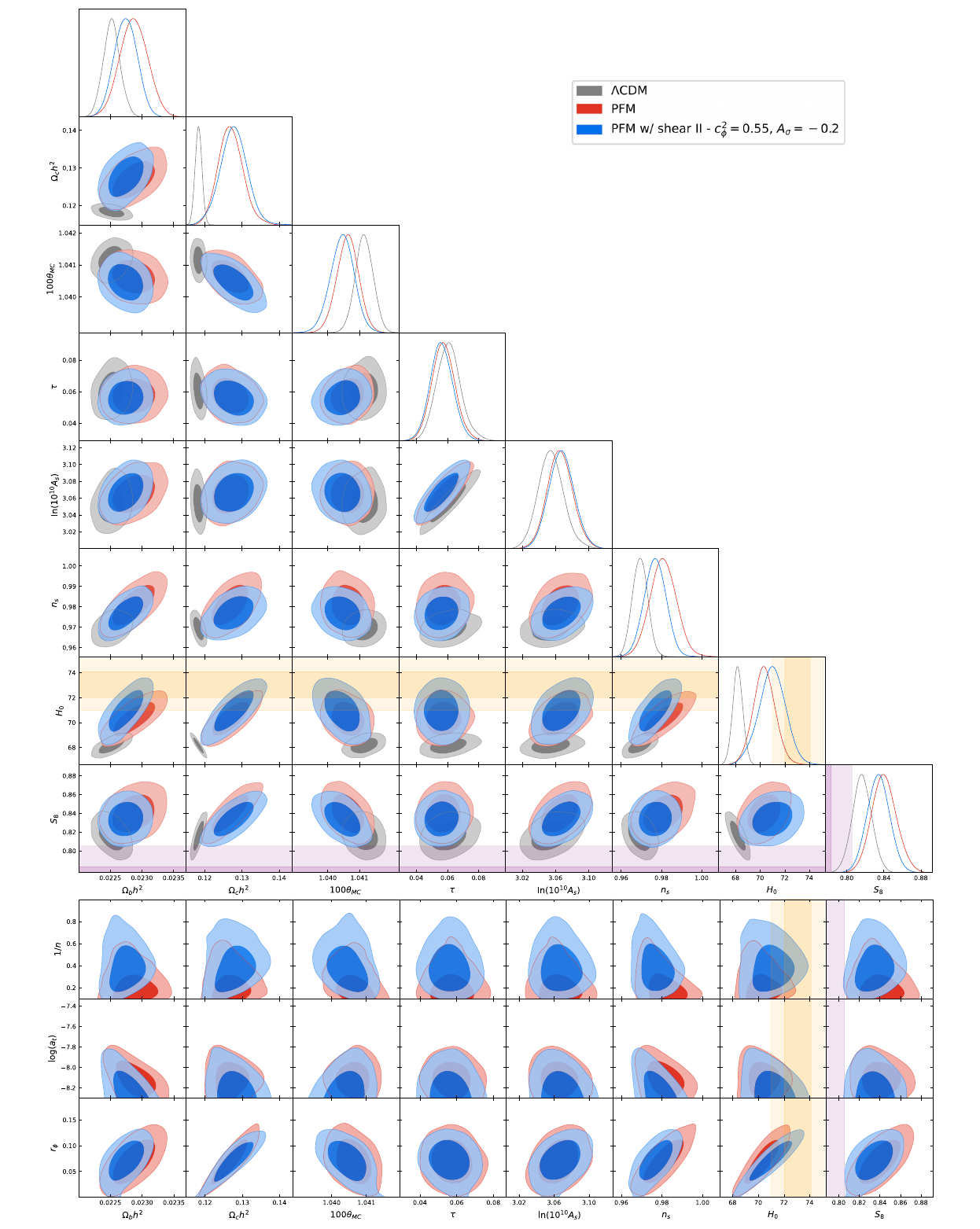}
     \caption{
     Top: Posterior distributions of the standard model parameters for the $\Lambda$CDM model (gray), the PFM with $c_\phi^2=1$ and $A_\sigma=0$ (red), and shear model II with $c_\phi^2=0.55$ and $A_\sigma=-0.2$ (blue). Bottom: Posterior distributions of the standard model parameters vs the background PFM parameters. The darker inner (lighter outer) regions correspond to $1\sigma$($2\sigma$) confidence intervals. The SH0ES Collaboration measurement of $H_0=73.04\pm1.04$ km/s/Mpc and the KiDS-1000 weak lensing survey measurement of $S_8=0.759^{+0.024}_{-0.021}$ are shown in the orange and purple bands, respectively \cite{Riess:2021jrx,KiDS:2020suj}. Distributions are generated with the P18+BAO+R19+SN datasets.
     }
     \label{fig: shearII background full}
 \end{figure*}

Finally, to see the broader impact of this negative shear model we re-run our analysis to include sampling over the background model parameters $r_\phi$, $a_t$, and $n$. We hold the microphysics fixed with $c_\phi^2=0.55$ and $A_\sigma=-0.2$ and explore the effect of this microphysics scenario on the background EDE solution. Parameter constraints on this case are given in Table \ref{tab: shearII background constraints} with their posterior distributions shown in Fig.~\ref{fig: shearII background P18} for the P18 dataset, and Fig.~\ref{fig: shearII background full} for the combined P18+BAO+R19+SN datasets. 

As can be seen by comparing Tables \ref{tab: background constraints} and \ref{tab: shearII background constraints}, the constraints on the model parameters follow a similar trajectory for a standard EDE model and for this negative shear case.  As with the standard EDE fluid model presented in Table \ref{tab: background constraints}, \textit{Planck} data alone shows no preference for EDE with a best-fit EDE density fraction of $r_\phi=0.005$. Hence, there is no solution to the Hubble tension with a best-fit $H_0=67.52$ km/s/Mpc, and the constraint on $S_8$ is unchanged from $\Lambda$CDM with $S_8=0.837(0.834)\pm0.013$. 

As discussed in Sec.~\ref{sec: resolution to the H0 tension}, for a nonzero amount of EDE to be preferred, we must include a late-universe prior on $H_0$. Any solution to the cosmological tensions would preferably exist in \textit{Planck} data alone, without the need for external datasets to enforce parameter changes. This is why in our main analysis of these models we fix the background evolution and investigate the effect of EDE microphysics under the assumption of a non-negligible EDE density around matter-radiation equality, allowing us to exclusively use \textit{Planck} data in our analysis. 

Considering the full dataset in this extended parameter space we find $S_8=0.834(0.836)\pm0.012$ for this negative shear model shown in Table \ref{tab: shearII background constraints}. Compared to the standard EDE fluid model constraint of $S_8=0.840(8.41)\pm0.013$ on this same dataset, we still see preference for a lower value of $S_8$. Comparing these values to the cases with fixed background evolution (shear model II with $A_\sigma=-0.2$ and the baseline EDE case, respectively) discussed in the main text, we see good agreement between models in both cases. However, for the full dataset considered here, the $\Lambda$CDM constraint is lowered to $S_8=0.816(0.817)\pm0.010$, making the softening of the $S_8$ tension weaker, but still statistically relevant as the standard EDE constraint lies outside the $\Lambda$CDM 1-$\sigma$ error bars. 

\begin{table*}[t]
\begin{tabular}{| c | c | c |}
\hline \hline
 Parameter & P18 only & P18+BAO+R19+SN \\ \hline \hline
 $100 \omega_b$ & $2.237(2.239)^{+0.016}_{-0.018}$ & $2.274(2.228)\pm0.018$ \\ 
 $\omega_c$ &  $0.1224(0.1206)^{+0.0014}_{-0.0026}$ & $0.1278(0.1292)\pm0.0036$ \\ 
 $100 \theta_s$  & $1.04073(1.04080)\pm0.00033$ & $1.04046(1.04044)\pm0.00038$ \\
 $\tau$ & $0.0540(0.0569)\pm0.0073$ & $0.562(0.0547)\pm0.0073$ \\ 
 $\ln (10^{10} A_s)$ & $ 3.049(3.049)^{+0.014}_{-0.015}$ & $3.067(3.070)\pm0.015$ \\ 
 $n_s$ & $0.9656(0.9657)^{+0.0046}_{-0.0054}$ & $0.9768(0.9791)\pm0.0054$ \\
 $1/n$ & $<0.600(0.331)$ & $0.397(0.377)^{+0.098}_{-0.180}$ \\
 $r_\phi$ & $<0.0196(0.0052)$ & $0.072(0.083)\pm0.026$ \\
 $a_t \times 10^4$ & $<4.35(2.59)$ & $<2.98(2.70)$ \\
 \hline
 $H_0$ [km/s/Mpc] & $67.97(67.52)^{+0.61}_{-1.10}$ & $71.0(71.50)\pm1.1$ \\ 
 $S_8$ & $0.837(0.834)\pm0.013$ & $0.834(0.836)\pm0.012$ \\
 \hline
 Total $\chi^2_\text{min}$ & 1013.77 & 2061.43 \\
 $\Delta \chi^2_\text{min}$ & -0.32 & -11.94 \\
 \hline \hline 
\end{tabular}
\caption{\label{tab: shearII background constraints} The mean(best-fit) $\pm1\sigma$ error on the cosmological parameters for the PFM w/ shear model II for the case of $c_\phi^2=0.55$ and $A_\sigma=-0.2$, with sampling over the background PFM parameters. }
\end{table*}

\bibliography{main}

\end{document}